\documentclass[10pt]{article}
\usepackage{a4}
\usepackage{color}
\usepackage[english]{babel}
\usepackage{epsfig}
\usepackage[T2A]{fontenc}
\usepackage{amsfonts}
\usepackage{amssymb}
\newcommand\mod{{~\rm mod~}}
\newcommand\XOR{{~\rm XOR~}}
\newcommand\OR{{~\rm OR~}}
\newcommand\AND{{~\rm AND~}}
\newcommand\determ{{~\rm det~}}
\newcommand{\PRNG}[1]{{\sf #1}}
\newcommand{\MCS}[1]{{\bf #1}}
\newcommand{\TEST}[1]{{\it #1}}
\newcommand{\reg}[1]{{\sf #1}}

\begin{document}
\title{Pseudo-random number generators for Monte Carlo simulations on Graphics Processing Units}
\author{Vadim Demchik\thanks{\tt E-mail: vadimdi@yahoo.com}
~\\~ {\small \sl{Dnepropetrovsk National University,
Dnepropetrovsk, Ukraine}}} \maketitle

\begin{abstract}
Basic uniform pseudo-random number generators are implemented on
ATI Graphics Processing Units (GPU). The performance results of
the realized generators (multiplicative linear congruential
(\PRNG{GGL}), XOR-shift (\PRNG{XOR128}), \PRNG{RANECU},
\PRNG{RANMAR}, \PRNG{RANLUX} and Mersenne Twister
(\PRNG{MT19937})) on CPU and GPU are discussed. The obtained
speed-up factor is hundreds of times in comparison with CPU.
\PRNG{RANLUX} generator is found to be the most appropriate for
using on GPU in Monte Carlo simulations. The brief review of the
pseudo-random number generators used in modern software packages
for Monte Carlo simulations in high-energy physics is present.
\end{abstract}

{\it Keywords:} Monte Carlo simulations, GPGPU, pseudo-random
number generators, performance

\section{Introduction}
The development of the General Purpose computing on Graphics
Processing Unit ({\it GPGPU}) technology has opened recently a new
cheap alternative to supercomputers and cluster systems for
researchers. The primary role of the GPGPU technology promotion
justly belongs to nVidia company which is the one of the two main
GPU hardware developers. nVidia motto is ``One Researcher, One
Supercomputer''. It fully reflects the tendencies of the
computational systems present-day market.  Moreover, the systems
which providing the GPGPU technology become the integral parts of
the contemporary supercomputers. In particular, while assembling
new supercomputer TH-1 with peak productivity 1.206 petaflops
which is equipped with the ATI GPUs {\textsf{HD 4870X2}} on every
node, China became the third country to build a petaflop
supercomputer (after USA and Germany) \cite{Top500}.

Pseudo-random numbers generation algorithms are key components in
most modern packages for researches and result depends on their
characteristics. The package should provide the actual
investigation of the physics or the other simulated problems, but
not to explore the behavior of the generator itself. Thus, while
choosing generator, one should consider not only its performance
productivity but also its statistical properties. Checking the key
results with generator of a different class than used one is also
very important \cite{Coddington:1997pc}. ``Pseudo-random'' term
actually means that for such values always exist an algorithm
which may reproduce the whole sequence. The values, produced by
some physical process, cannot be considered as random in a general
sense, because even if we cannot predict the sequence of such
numbers, it does not mean that there is no algorithm to produce
them \cite{Marsaglia:2003rn}.

The main aim of the present paper is to show the possibility to
adapt for GPU the most commonly used pseudo-random number
generators ({\it PRNG}) as the main components of the software
packages for Monte Carlo ({\it MC}) simulations. An essential
distinction of the GPU architecture from the Central Processing
Unit ({\it CPU}) architecture causes new difficulties as well as
new optimization capabilities. The question about the performances
of the different PRNGs on GPU platform is arisen in this
connection. The performance results and the difference between
PRNGs performances on CPU and GPU will be shown in the paper.

We will restrict our investigation only to uniform PRNGs. In
particular some generators were already investigated for nVidia
GPUs in \cite{Thomas:2009ug, Langdon:2009fh}. But studied
generators are not the most frequently used ones in MC
simulations.

In the present paper we will not refer the question of the
random-number generator testing (see for example
\cite{Vattulainen:1994nt} and references therein) and will content
only with implementation of existing generators on GPU.

The paper is organized as follows. Section \ref{PRNGMC} contains
the brief list of the software packages for MC simulations in
high-energy physics with showing PRNGs which are used in the
corresponding package. The details of the GPU implementation and
the code listings of the realized PRNGs are described in section
\ref{GPUimplementation}. The performance results and discussion
are collected in section \ref{resultsdiscuss}. In section
\ref{conclusions} we draw our conclusions. Finally, appendix
includes some theoretical background for the realized PRNG
algorithms.

\section{PRNGs in Monte Carlo simulations}\label{PRNGMC}
In this section we present the list of PRNGs which are employed in
nowadays MC simulations.

To choose a generator which will be used in MC simulations, it is
not enough to consider only its period and algorithm output. The
generator \PRNG{R250} was widely used due to its simplicity and
long period, but it was found that this generator has the
essential statistical defects which make impossible to use it in
modern simulations (see \cite{Janke:2002} and reference therein).
Apart from general statistical tests like \TEST{DIEHARD Battery of
Tests of Randomness} and \TEST{Crush}, a generator has to pass
empirical test in real conditions. That is why, while developing
MC application, usually only generators with minimal statistical
influence for the present MC simulations are used. For example,
European Organization for Nuclear Research ({\it CERN}) recommends
using three PRNGs: \PRNG{RANMAR} (V113), \PRNG{RANECU} (V114) and
\PRNG{RANLUX} (V115).

Below is the incomplete list of the software packages for MC
simulations in high-energy physics and the PRNG which is used in
corresponding package.
\begin{itemize}
    \item \MCS{FermiQCD}: is an open C++ library
    for development of parallel Lattice Quantum Field Theory
    computations \cite{DiPierro:2005qx,FermiQCD}. It uses
    floating-point version of \PRNG{RANMAR} generator as default PRNG.
    \item \MCS{UKQCD}: By indirect information UKQCD Collaboration also uses
    \PRNG{RANMAR} PRNG \cite{DiPierro:2000nt} in its software.
    \item \MCS{MILC}: is an open code of high performance research
    software written in C for doing $SU(3)$ lattice gauge theory
    simulations on several different (MIMD) parallel computers \cite{MILC}.
    \MCS{MILC} uses own XOR of a 127-bit feedback shift register and
    a 32 bit integer congruential generator. Each node or site uses a different
    multiplier in the congruential generator and different initial state in the
    shift register. So, all nodes are generating different sequences of
    numbers.
\begin{eqnarray}
t=(\left((X_{n-5}\gg 7)\OR(X_{n-6}\ll 17)\right)\XOR\\\nonumber
  \left((X_{n-4}\gg 1)\OR(X_{n-5}\ll 23)\right))\AND (2^{24}-1),\\\nonumber
  X_{n-6}=X_{n-5},~~X_{n-5}=X_{n-4},~~X_{n-4}=X_{n-3},~~X_{n-2}=X_{n-1},\\\nonumber
  X_{n-1}=X_{n},~~X_{n}=t,~~Y_{n}=a_jY_{n-1}+c,\\\nonumber
  z_{n}=(t\XOR ((Y_{n}\gg 8)\AND (2^{24}-1)))/2^{24}
\end{eqnarray}
    \item \MCS{CPS}: The Columbia Physics System is a large set of codes for
    lattice QCD simulations \cite{CPS}. \PRNG{RAN3} is used.
    \item \MCS{SZIN}: is the open-source software system supports
    data-parallel programming constructs for lattice field theory
    and in particular lattice QCD \cite{SZIN}. \PRNG{RANLUX} is used
    in \MCS{SZIN} packet.
    \item \MCS{ISAJet}: is a Monte Carlo event generator for $pp$, $p\bar{p}$,
    $e^+e^-$ interactions \cite{Paige:2003mg,ISAJet}. \PRNG{RANLUX}
    is incorporated into \MCS{ISAJet}.
    \item \MCS{GEANT4}: is a toolkit for the simulation of the passage of
    particles through matter \cite{Geant4}. \MCS{GEANT4} uses the \MCS{HEPRandom}
    module \cite{HEPRandom} to generate pseudo-random numbers which includes 12 different
    random engines (\PRNG{RANMAR}, \PRNG{RANECU}, \PRNG{DRAND48},
    \PRNG{RANLUX}, etc) now.
    \item \MCS{PYTHIA}: is a program for the generation of high-energy physics
    events \cite{Sjostrand:2006za,PYTHIA}. \PRNG{RANMAR} is used in \MCS{PYTHIA}
    as an internal PRNG.
    \item \MCS{HERWIG}: is a Monte Carlo package for simulating hadron emission
    reactions with interfering gluons \cite{Corcella:2000bw,HERWIG}.
    \PRNG{RANECU} is used.
    \item \MCS{CompHEP}: a package for evaluation of Feynman diagrams,
    integration over multi-particle phase space and event generation
    \cite{Pukhov:1999gg,CompHEP}. \PRNG{DRAND48} is used in \MCS{CompHEP}.
    \item \MCS{MC@NLO}: is a Fortran package to implement the scheme for
    combining a Monte Carlo event generator with Next-to-Leading-Order
    calculations of rates for QCD processes \cite{Frixione:2002ik,MCNLO}.
    \PRNG{GGL} is used.
    \item \MCS{SHERPA}: is a Monte Carlo event generator for the simulation
    of high-energy reactions of particles in lepton-lepton, lepton-photon,
    photon-photon and hadron-hadron collisions
    \cite{Gleisberg:2003xi,SHERPA}. \PRNG{RANMAR} is used.
    \item \MCS{Chroma}: is a software system for lattice QCD calculations
    \cite{Edwards:2004sx,Chroma}.  In \MCS{Chroma} slightly modified
    linear congruential generator \PRNG{RANNYU} is implemented --
    LCG($a=31167285$,$m=2^{48}$).
    \item \MCS{GENIE}: is a neutrino event generator for experimental
    physics community \cite{Andreopoulos:2009rq,GENIE}. \PRNG{MT19937} is used.
    \item \MCS{ALPGEN}: is an event generator for hard multiparton processes
    in hadronic collisions \cite{Mangano:2002ea,ALPGEN}. \PRNG{RANECU} is used.
\end{itemize}

So, the most commonly used generators are \PRNG{RANMAR},
\PRNG{RANLUX}, \PRNG{RANECU} and several variations of a linear
congruential generator (\PRNG{DRAND48}, \PRNG{RAN3}, \PRNG{GGL},
\PRNG{RANYU}). Also, most new packages began to include the
generator \PRNG{MT19937} as internal PRNG. Hence, it is reasonable
to implement the generators on GPU which are used in real MC
simulations.

\section{GPU implementation}\label{GPUimplementation}
In this section we give the base ideas for PRNG implementation on
GPU and describe the source codes of the following PRNGs
realizations on ATI GPUs: \PRNG{GGL}, \PRNG{XOR128},
\PRNG{RANECU}, \PRNG{RANMAR}, \PRNG{RANLUX} and \PRNG{MT19937}.

We use ATI Stream SDK \cite{ATIsdk} for the realization of the
PRNGs on ATI GPU as software environment. ATI CAL allows using ATI
GPU hardware in the most effective way \cite{Demchik:2009ni}. That
is why all PRNGs realizations presented below are made on ATI
Intermediate Language ({\it IL}) \cite{ATIIL}, and not on higher
level, for example OpenCL or Brook+.

Three different ATI video cards are used to check the algorithm
efficiency -- ATI Radeon {\textsf{HD 5850}}, {\textsf{HD 4870}}
and {\textsf{HD 4850}}. The essential for GPGPU-applications
parameters about some ATI's video cards are presented in Table
\ref{GPUinfo}.
All cards are equipped with the GDDR5 memory except {\textsf{HD
4850}} which contains GDDR3 memory (this fact was marked with the
asterisk). GDDR5 memory possesses a quadruple effective data
transfer rate relative to its physical clock rate, instead of
double as with GDDR3 memory. It is necessary to note that
{\textsf{HD 4870X2}} and {\textsf{HD 5970}} are two-core cards.
For our purposes at program level they are equivalent to two
devices installed in the system.

\begin{table}
\caption{General information about some ATI's video cards
\cite{wikiATI}.}\label{GPUinfo}\small
\begin{center}
\begin{tabular}{|l|c|c|c|c|c|c|}\hline
 \rule{0pt}{10pt} Model & Stream & Core & Memory & Bandwidth & Bus width &
 Tflops\\
 & cores & (MHz) & (MHz) & (GB/s) & (bit) & (peak)\\\hline
 \rule{0pt}{10pt} HD 4850 & 800 & 625 & $993^*$ & 64 & 256 & 1.0\\\hline
 \rule{0pt}{10pt} HD 4870 & 800 & 750 & 900 & 115.2 & 256 & 1.2\\\hline
 \rule{0pt}{10pt} HD 4870X2 & $2\times 800$ & $2\times 750$ & 900 & $2\times 115.2$ & $2\times 256$ & 2.4\\\hline
 \rule{0pt}{10pt} HD 5850 & 1440 & 725 & 1000 & 128 & 256 & 2.1\\\hline
 \rule{0pt}{10pt} HD 5870 & 1600 & 850 & 1200 & 153.6 & 256 & 2.7\\\hline
 \rule{0pt}{10pt} HD 5970 & $2\times 1600$ & $2 \times 725$ & 1000 & $2\times 128$ & $2\times 256$ & 4.6\\\hline
\end{tabular}
\end{center}
\end{table}

\subsection{General implementation scheme}\label{SubSectionGeneralScheme}
To design the GPU-applications it must be accounted the following
main features of hardware architecture:
\begin{itemize}
    \item each general-purpose register and memory cell has the four 32-bit components
          that are designated as $.x$, $.y$, $.z$ and $.w$;
    \item floating point operations are more productive on GPU than integer
          operations (in compare with CPUs);
    \item double precision floating point operations are the
          slowest on GPU;
    \item ATI GPU can perform up to five operations on each VLIW processor
          simultaneously.
\end{itemize}
Computing programs working for GPU are known as kernels. All
kernels are run by the host program. Each kernel must perform a
complete operation, because of ATI Stream SDK does not support the
execution of a kernel from another kernel (except OpenCL
applications).

We offer to use PRNGs directly in MC procedures as the separate
subroutines. Undoubtedly such scheme allows the possibility to
keep generated random numbers in GPU-memory, for example, for
further usage by other procedures, as it is usually made in
CPU-simulations. However, it seems to be more effective to
``virtualize'' produced by PRNG pseudo-random numbers as it allows
to eliminate unnecessary additional read-write operations to the
GPU-memory for the random numbers as well as to decrease the
GPU-memory consumption of the application.

In order to accelerate GPU-applications work, we recommend to keep
all data needed for kernel operations directly in GPU-memory,
because memory operations are a bottleneck of GPU.

For MC simulations on GPU it is convenient to produce four random
numbers through one PRNG pass which is closely related to GPU
memory architecture. Under such conditions one can use GPU
performance in the most efficient way. As a rule more than one
random number are used in MC simulations for updating (for
example, one for update proposition and one for probability). So,
it seems to be natural to generate the corresponding number of the
pseudo-random numbers for further purposes.

\begin{figure}
\begin{center}
\resizebox{0.6\textwidth}{!}{\includegraphics{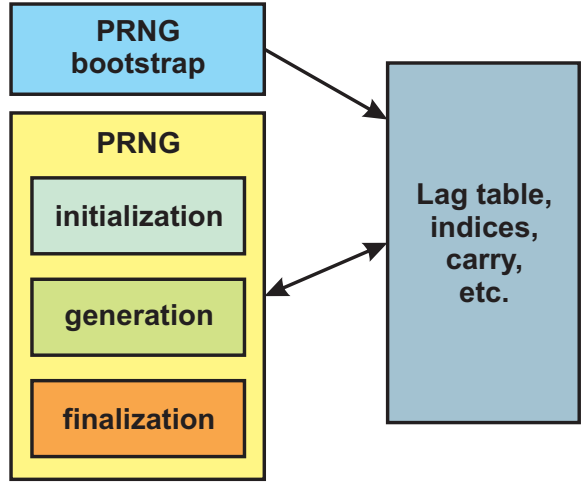}}
\caption{PRNG structure scheme} \label{fig:1}
\end{center}
\end{figure}

The common structure scheme of PRNGs is shown in Figure
\ref{fig:1}. All implemented PRNGs are divided into 3 phases:
\begin{itemize}
    \item {\it initialization:} loading and preparing the
    previous state of PRNG and all preparing operations for PRNG;
    \item {\it random number generation:} actually all PRNG operations
    which generate pseudo-random numbers;
    \item {\it finalization:} storing the final state of PRNG for
    the next run.
\end{itemize}
It allows to make the best use of PRNG, as it is needed a few more
random numbers. For example, at multihit updating method the
several pseudo-random numbers are required per one pass of the
updating procedure. The initialization subroutine of the PRNG is
called on the start of the updating procedure. During the work the
updating procedure may call the PRNG generation subroutine many
times. And finally, at the end it must call the finalization
subroutine of the PRNG.

All PRNG lag tables are stored directly in GPU-memory and it
avoids needless transfer the data between CPU and GPU memory. Each
instance of the PRNG uses its own lag table to parallelize the
process. The universal method of the paralleling is used in all
realized PRNGs -- using the independent sequences. To be exact,
every running instance of PRNG obtains its own index $J=0, 1,
\ldots, J_{max}$. The lowest bits of $J$ serve to identify the lag
table for PRNG. Another possible scheme which may be used is the
selection of the lag table through the modulo operation.

We use a little bigger number of actual PRNG instances than the
number of the stream cores in particular GPU to avoid possible
collisions among threads. For example, top one-unit ATI GPU card
{\textsf{HD 5870}} has 1600 stream cores (see Table
\ref{GPUinfo}), while 13 lowest bits were used to identify the lag
table for PRNG. It corresponds to the 8192 parallel instances of
the PRNG.

Almost all generators require the initialization procedure to
bootstrap the lag table (three of presented here generators,
\PRNG{RANMAR}, \PRNG{RANLUX} and \PRNG{MT19937}). This procedure
is realized directly on the GPU unit, not with the help of the CPU
with further transfer the lag table into GPU-memory. We do not
show lag table initialization procedure here to save the space in
the article.

For convenience and performance purposes all the kernels have been
precompiled to replace all constants \reg{\%VariableName\%} with
the corresponding values. It allows to avoid the constant buffer
using and put the runtime constant parameters directly in
assembler code. Under \reg{\%VariableName\%} the value of the
respecting variable will be implied in source codes below.
Sometimes before the \reg{VariableName} it will stay the prefix
\reg{i} (like \reg{\%iVariableName\%}) which will show the decimal
format instead of the default hexadecimal format. Decimal format
is needed to specify the relative offsets in global memory
operations.

Note that each of the presented generators can be easily modified
for the generation of the pseudo-random numbers with double
precision. All presented generators are either 24-bit, or 32-bit,
while for number representation with double precision 64 bits are
required. Thus, for correct generation pseudo-random numbers with
double precision one should use several numbers with single
precision, but not only covert a number with the single precision
into the double precision number by the regular conversion command
(it considerably decreases the quantity of possible realization).

\subsection{GGL}
\PRNG{GGL} is one of the simplest and computational ``light''
portable PRNG which could be implemented on GPU. \PRNG{GGL} has
been studied by Langdon on nVidia Tesla T10P with CUDA SDK
\cite{Langdon:2009fh}. The obtained peak performance of the PRNG
is about $2\times 10^9$ pseudo-random numbers per second. It is
obvious that while period $P_{\PRNG{GGL}}\leq 2147483646$ and its
run in 1024-threads and output $10^{9}$ pseudo-random numbers per
second the \PRNG{GGL} period could be exhausted in about $0.002$
second. So, \PRNG{GGL} are realized here just to check the
performance of the pseudo-random number production on ATI GPUs on
a very simple generator which requires to store only one seed
value per PRNG instance.

\begin{figure}
\begin{center}
\resizebox{1.0\textwidth}{!}{\includegraphics{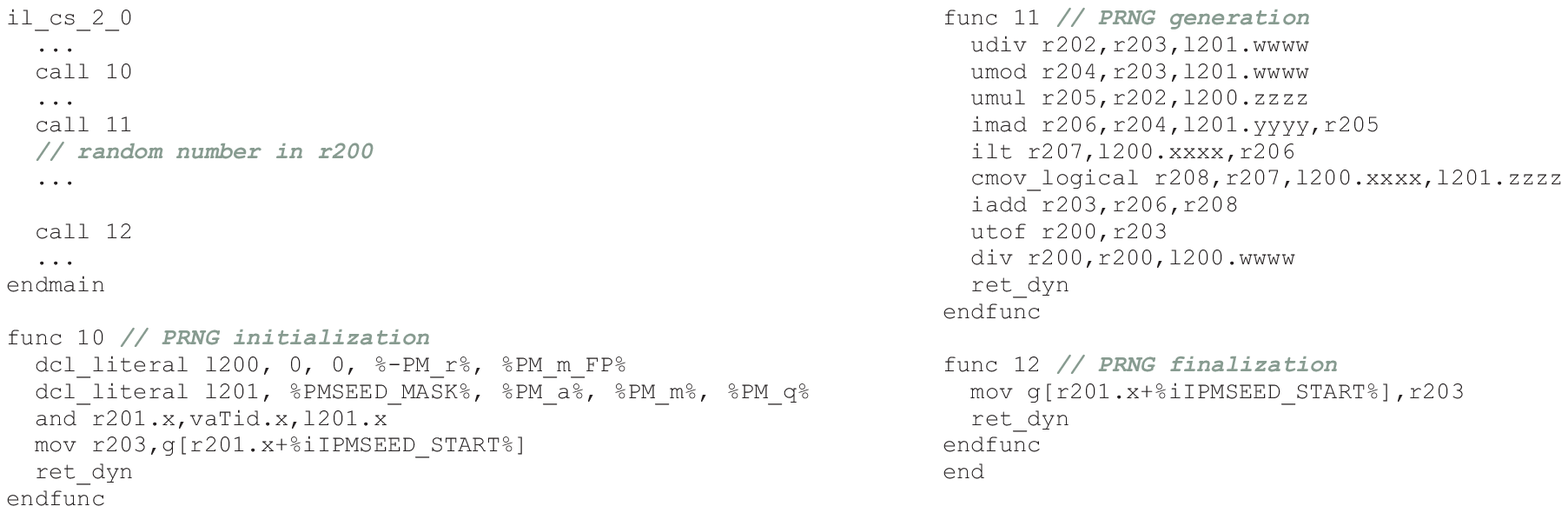}}
\caption{Source code of \PRNG{GGL} PRNG} \label{fig:PM}
\end{center}
\end{figure}

Park and Miller have published the four Pascal implementations of
the \PRNG{GGL} \cite{Park:1988rn}, for integer and real arithmetic
(one direct scheme and one avoiding the overflow). For
implementation we chose the {\it ``Integer version 2''} of the
\PRNG{GGL} \cite{Park:1988rn}. The ATI IL source code of
\PRNG{GGL} PRNG is presented in Figure \ref{fig:PM}. There are
three subroutines 10, 11 and 12 after the main module. Subroutines
10 and 12 are called only once and perform the initialization and
the finalization of the PRNG, correspondingly. Subroutine 11
produces four-component pseudo-random numbers in register
\reg{R200} per one pass. This program scheme fully corresponds to
the basic scheme, described in subsection
\ref{SubSectionGeneralScheme}, and hereinafter will be used for
the rest of generators presented in this work.

The used variables are:
\begin{eqnarray}\nonumber
 \reg{PM\_m}=2147483647, && \reg{PM\_m\_FP}=2147483647.0,\\\nonumber
 \reg{-PM\_r}=-2836, && \reg{PM\_a}=16807,\\\nonumber
 \reg{PM\_q}=127773, &&
\end{eqnarray}
\reg{PMSEED\_MASK} specifies the PRNG instance and
\reg{iIPMSEED\_START} is the offset of the lag table in the global
buffer which is prepared by the host program.

The parallelization scheme with separate sequences is implemented
here -- every GPU thread produces four separate pseudo-random
sequences (by every slot of general purpose register \reg{R200}).
So, the whole number of the pseudo-random sequences is the
quadruple number of the threads to be run. The threads must be
initialized carefully to reach the maximal period length of the
PRNG and avoid the sequences overlapping.

Generator requires to keep only one seed-value per thread for work
which is equal to choosing only one four-component cell per thread
in global memory which is read and kept only once per PRNG cycle.
Thus, for 4096-thread run (or 16384 subthreads) the size of the
lag table will be 64kB. Initial seeds are filled with the host
program and transferred to GPU memory before the first run of the
\PRNG{GGL}. Seed values must be exactly chosen to rich the maximal
period of PRNG. But in our case the performance of the generator
is the main aim, so all the seeds are selected randomly, because
the generator exhausted whole the period for very short time and
it is not so important when and where the overlapping of the
sequences will began.

The other LCGs could be easily realized on the given example of
the PRNG by substituting the corresponding LCG parameters.
\PRNG{GGL} generator possesses very good performance, as it will
be shown in the next section. Nevertheless, in spite of the PRNG
performance, it would not be used for practical purposes due to
very short period.

\subsection{XOR128}
Next PRNG implemented on ATI GPU is \PRNG{XOR128}, a very fast
generator with much better statistical properties and considerably
longer period than \PRNG{GGL}. The distinctive feature of this
PRNG is the usage of the four-component 32-bit values to produce
the sequence which exactly corresponds to bit-capacity of GPU
memory.

\begin{figure}
\begin{center}
\resizebox{1.0\textwidth}{!}{\includegraphics{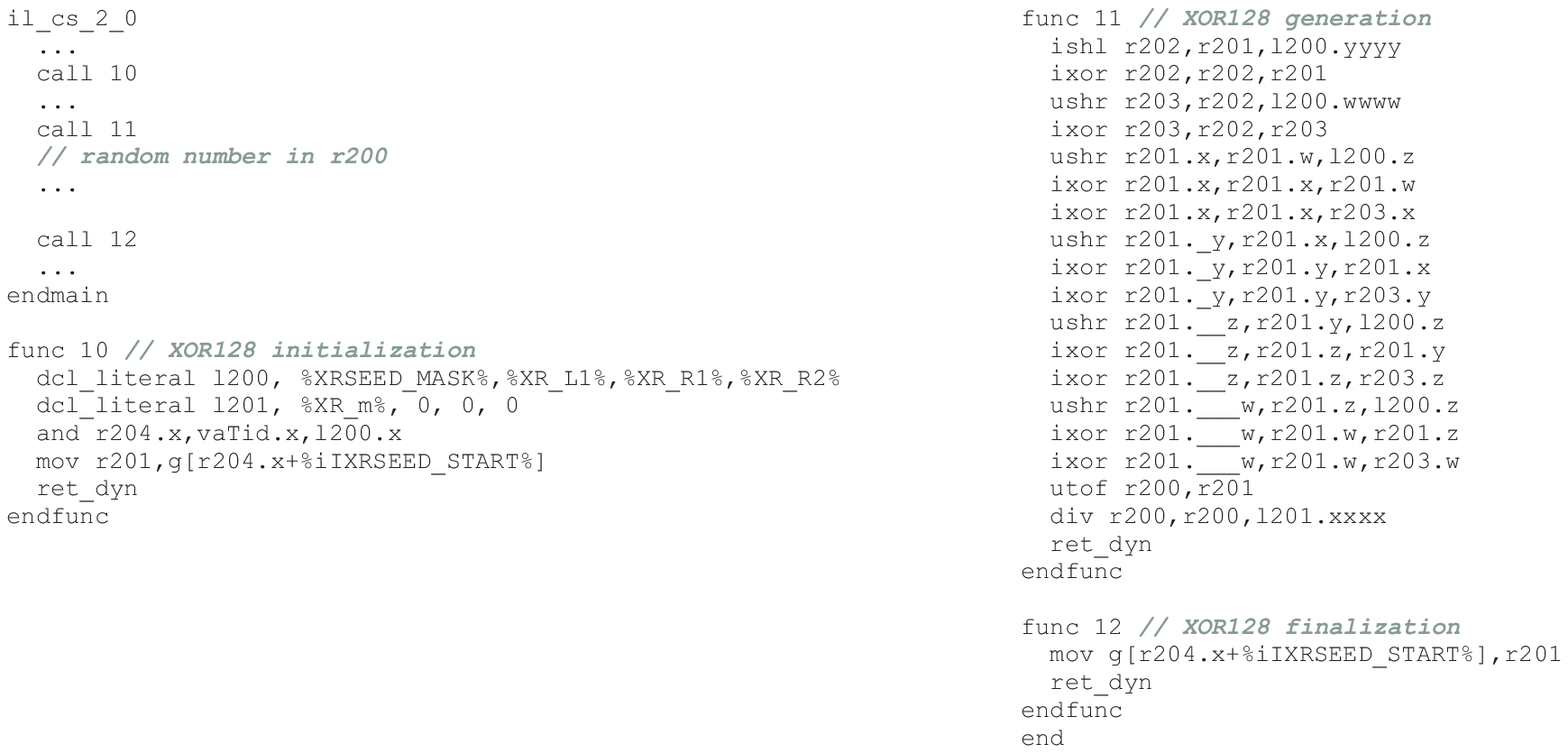}}
\caption{Source code of \PRNG{XOR128} PRNG} \label{fig:XOR128}
\end{center}
\end{figure}

The ATI IL source code of \PRNG{XOR128} is shown in Figure
\ref{fig:XOR128}. The following variables are used
\begin{eqnarray}\nonumber
 \reg{XR\_L1}=11, && \reg{XR\_R1}=19,\\\nonumber
 \reg{XR\_R2}=8, && \reg{XR\_m}=4294967295.0,
\end{eqnarray}
\reg{XRSEED\_MASK} specifies the PRNG instance,
\reg{iIXRMSEED\_START} is the offset of the lag table in global
buffer which is prepared by host program.

The number of sequences produced by \PRNG{XOR128} corresponds to
the number of threads. One thread generates four successive
pseudo-random numbers from one sequence in every components of the
general purpose register \reg{R200}. Marsaglia's algorithm
\cite{Marsaglia:2003xr} is slightly modified to parallelize
sequence production into four subthreads -- four items of the lag
table are composed in one four-component memory cell and are
produced in one pass. Unfortunately, it is impossible to avoid
recursion without making the algorithm more complicated. Therefore
among the four-component operations in the source code there are
single-slot operations which are partially parallelized further by
the IL compiler.

Except the final integer-to-float conversion operations in the
algorithm, there are only bit shift and exclusive-OR operations
which are ``computationally light'' GPU operations in compare with
integer operations. Along with few memory operations (one read and
one written operations per run), it also increases the performance
of PRNG on GPU.

The \PRNG{XOR128} generator has a period large enough to be used
on GPU. For 2048-threads run and average performance about
$10^{10}$ samples per second it could be exhausted in about
$10^{17}$ years only. And only strong criticism of L'Ecuyer
\cite{Lecyuer:2005xs} does not to allow use \PRNG{XOR128} as
standard PRNG on GPU.

Generator requires keeping four 32-bit integers per thread. In the
present realization PRNG produces four sequential pseudo-random
numbers per thread which allows reserving only one 4-component
cell per thread in global memory. This 4-component seed is read
and written only once per PRNG cycle. The size of the
\PRNG{XOR128} lag table is the same to \PRNG{GGL}, i.e. for
4096-thread run it takes the 64kB of the GPU memory.

\subsection{RANECU}
Another high-performed PRNG realized on ATI GPU is \PRNG{RANECU}.
This generator is recommended by CERN and used in some software
packages, listed in the previous section. Relatively long period
and quite good statistical properties make \PRNG{RANECU}
reasonably attractive generator for small tasks.

\begin{figure}
\begin{center}
\resizebox{1.0\textwidth}{!}{\includegraphics{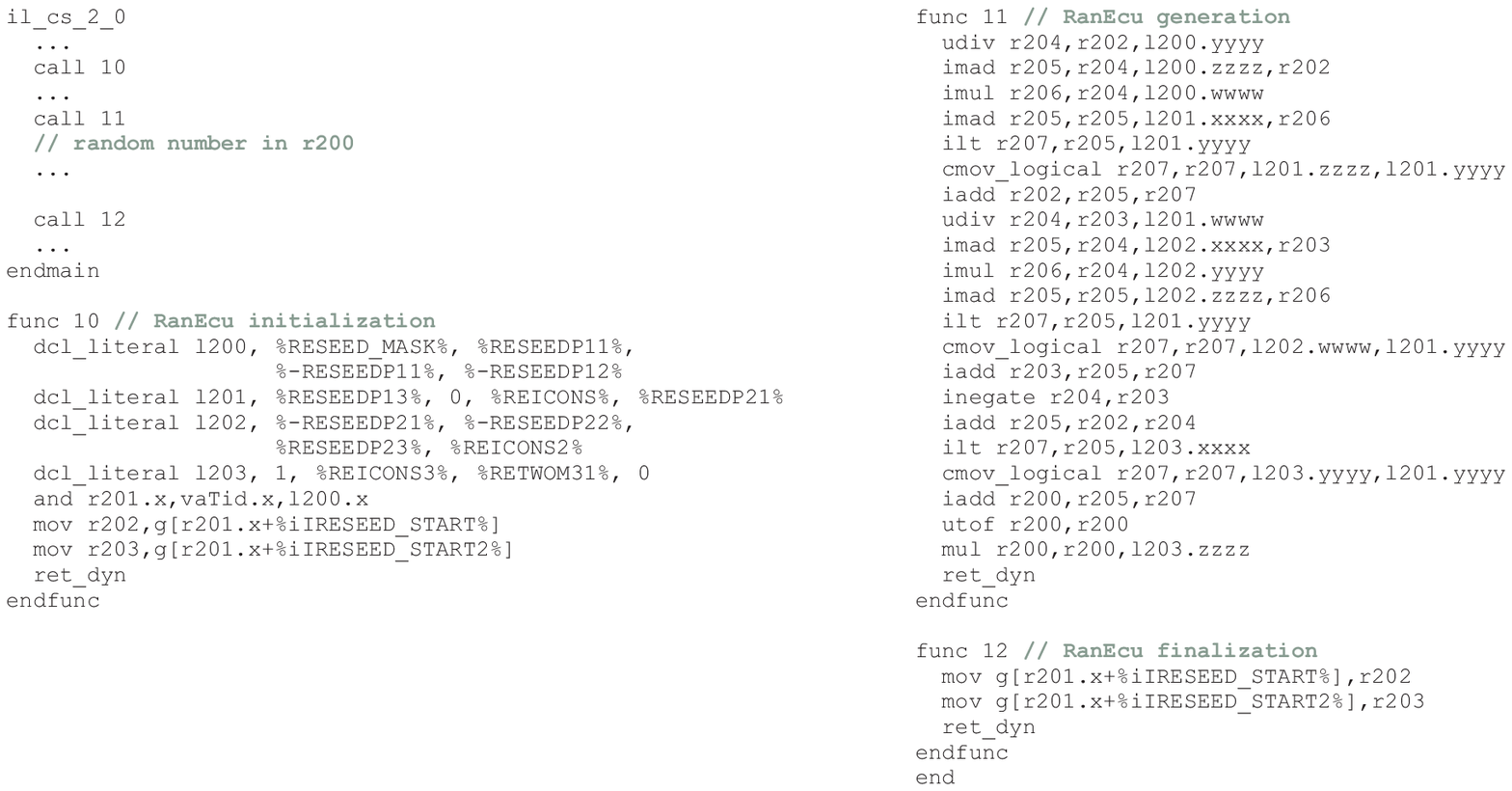}}
\caption{Source code of \PRNG{RANECU} PRNG} \label{fig:RANECU}
\end{center}
\end{figure}

The scheme like in the case of \PRNG{GGL} generator is implemented
here: each thread produces the four independent sequences which
are composed into the four components of the general purpose
register \reg{R200}. For \PRNG{RANECU} run in 1024-threads and
output $5\times 10^{9}$ pseudo-random numbers per second the
\PRNG{RANECU} period could be exhausted in about 31 hours. Despite
this fact the generator is still acceptable for wide range of MC
simulation tasks.

The ATI IL source code of the \PRNG{RANECU} PRNG is presented in
Figure \ref{fig:RANECU}. The following variables are used
\begin{eqnarray}\nonumber
 \reg{RESEEDP11}=53668, && \reg{-RESEEDP11}=-53668,\\\nonumber
 \reg{-RESEEDP12}=-12211, && \reg{RESEEDP13}=40014,\\\nonumber
 \reg{RESEEDP21}=52774, && \reg{-RESEEDP21}=-52774,\\\nonumber
 \reg{-RESEEDP22}=-3791, && \reg{RESEEDP23}=40692,\\\nonumber
 \reg{REICONS}=2147483563, && \reg{REICONS2}=2147483399,\\\nonumber
 \reg{REICONS3}=2147483562, && \reg{RETWOM31}=1.0/2147483648.0,
\end{eqnarray}
\reg{iIRESEED\_START} and \reg{iIRESEED\_START2} are the offsets
of two tables of seeds in global buffer, \reg{RESEED\_MASK}
specifies the PRNG instance. In fact, the \PRNG{RANECU} lag table
is divided into two tables for convenience here. These tables are
prepared by the host program.

The generator requires keeping two integer seed-values per thread
which are grouped into two lag subtables. So for 4096-thread run
the size of the lag table is 128kB.

The generator kernel is made on integer arithmetic base which
slightly brings down the generator performance while using GPUs.
However, simplicity of the algorithm and small amount of the lag
table elements completely compensate this ``drawback''. On the
base of the present code, one can easily construct another MRG by
substitution of the corresponding parameters.

\subsection{RANMAR}
The next generator realized on ATI GPU is the 24-bit Marsaglia
PRNG \PRNG{RANMAR}. It was previously implemented on ATI GPU in
\cite{Demchik:2009ni} for Ising model and $SU(2)$ gluodynamics
simulations.

In contrast to previously presented PRNGs, \PRNG{RANMAR} has a
larger lag table which contains 97 elements. All these lag table
items must be prepared before the first working pass of the
generator. Of course, the lag table may be directly initialized by
the user, but it is not convenient in practice. So, the
\PRNG{RANMAR} lag table is initialized by stand-alone procedure
RMARIN, proposed by James \cite{James:1988vf}. In this procedure,
the whole lag table is initialized on the base of only two given
5-digit integers, each set of which causes an independent sequence
of the sufficient length for an entire calculation. The seed
variables can have values between $0$ and $31328$ for the first
variable and $0$ and $30081$ for the second variable,
respectively. \PRNG{RANMAR} can create, therefore, 900 million
independent subsequences for different initial seeds with each
subsequence having a length of about $10^{30}$ pseudo-random
numbers. This approach considerably reduces the number of the
possible generator states. Still it brings an important element of
the generator features -- easy division of sequences produced by
generator among PRNG instances without overlapping. Possible
sequences quantity (900 millions) at existing or developing
hardware is considered to be sufficient even in medium-term
perspective.

The generator consists of two parts:
\begin{itemize}
    \item kernel which produces the seed numbers on initial seed values;
          in fact this kernel is a replica of the RMARIN subroutine of
          the James' version \cite{James:1988vf};
    \item subroutines which directly produce the random numbers.
\end{itemize}
First part is executing only for initialization.

The floating-point version of the \PRNG{RANMAR} generator is
realized here which produces the pseudo-random directly in the
interval $[0;1)$. So, it is not needed to use slowest
integer-to-floating point converting operation.

\begin{figure}
\begin{center}
\resizebox{1.0\textwidth}{!}{\includegraphics{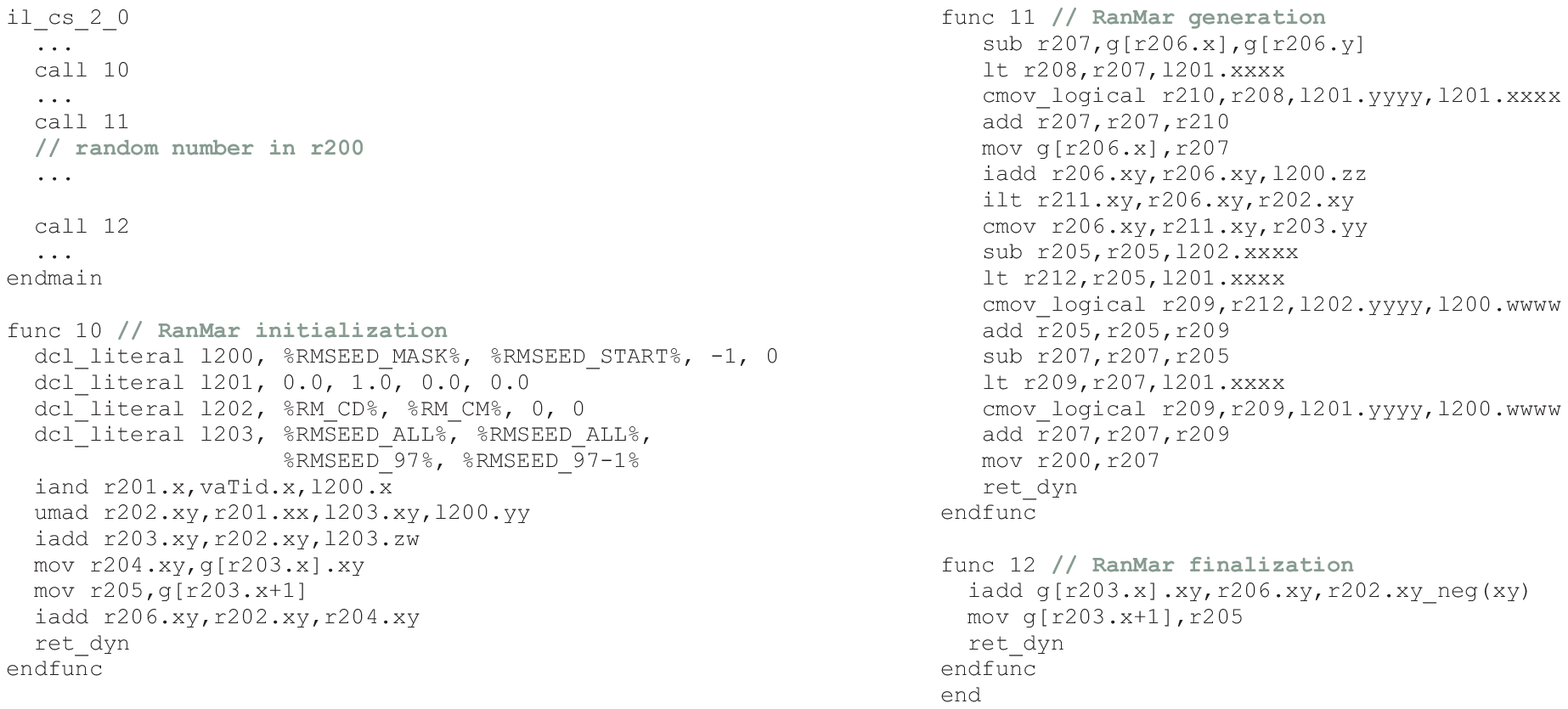}}
\caption{Source code of \PRNG{RANMAR} PRNG} \label{fig:RANMAR}
\end{center}
\end{figure}

Apart from 97 elements in the lag table each \PRNG{RANMAR}
instance must store previous value of arithmetic sequence and two
indices which are connected to each other. As in the case of
\PRNG{GGL} PRNG, every GPU thread produces four independent
sequences in the presented implementation of \PRNG{RANMAR}.
Obviously at such approach, it is necessary to keep only one pair
of indices for all four subthreads, because these subthreads are
executed out synchronously. So, it requires 2.5 memory cells be
read and written for one PRNG pass. The size of \PRNG{RANMAR} lag
table for 4096-thread run is about 6MB. Initially lag table is
prepared by stand-alone procedure RMARIN which is not shown here.

The ATI IL source code of \PRNG{RANMAR} PRNG is presented in
Figure \ref{fig:RANMAR}. The following variables are used
\begin{eqnarray}\nonumber
 \reg{RM\_CD}=7654321.0/16777216.0, && \reg{RM\_CM}=16777213.0/16777216.0,\\\nonumber
 \reg{RMSEED\_97}=97, && \reg{RMSEED\_97-1}=96,\\\nonumber
 \reg{RMSEED\_ALL}=99,
\end{eqnarray}
\reg{RMSEED\_START} is the offset of the lag table in global
buffer, \reg{RMSEED\_MASK} specifies the PRNG instance.

\subsection{RANLUX}\label{RANLUXimplementation}
Nowadays \PRNG{RANLUX} PRNG is one of the standard high-performed
generators for Monte-Carlo simulations. The statistical properties
of the generator are well-known. From the realization point of
view, distinctive feature of the generator is the necessity to
discard out groups of generated pseudo-random numbers after one
generation cycle. Omitted values quantity is determined by the
``luxury'' parameter. While implementation of the algorithm on the
central processing unit such discarding is ``virtual'', because
lag table fits in processor cache very well, as a rule. Still this
algorithm phase is very resource-intensive on GPU, because global
buffer is not a generally cached object. Thus it is obvious, that
the PRNG performance is strongly depend on this phase of algorithm
realization.

\begin{figure}
\begin{center}
\resizebox{1.0\textwidth}{!}{\includegraphics{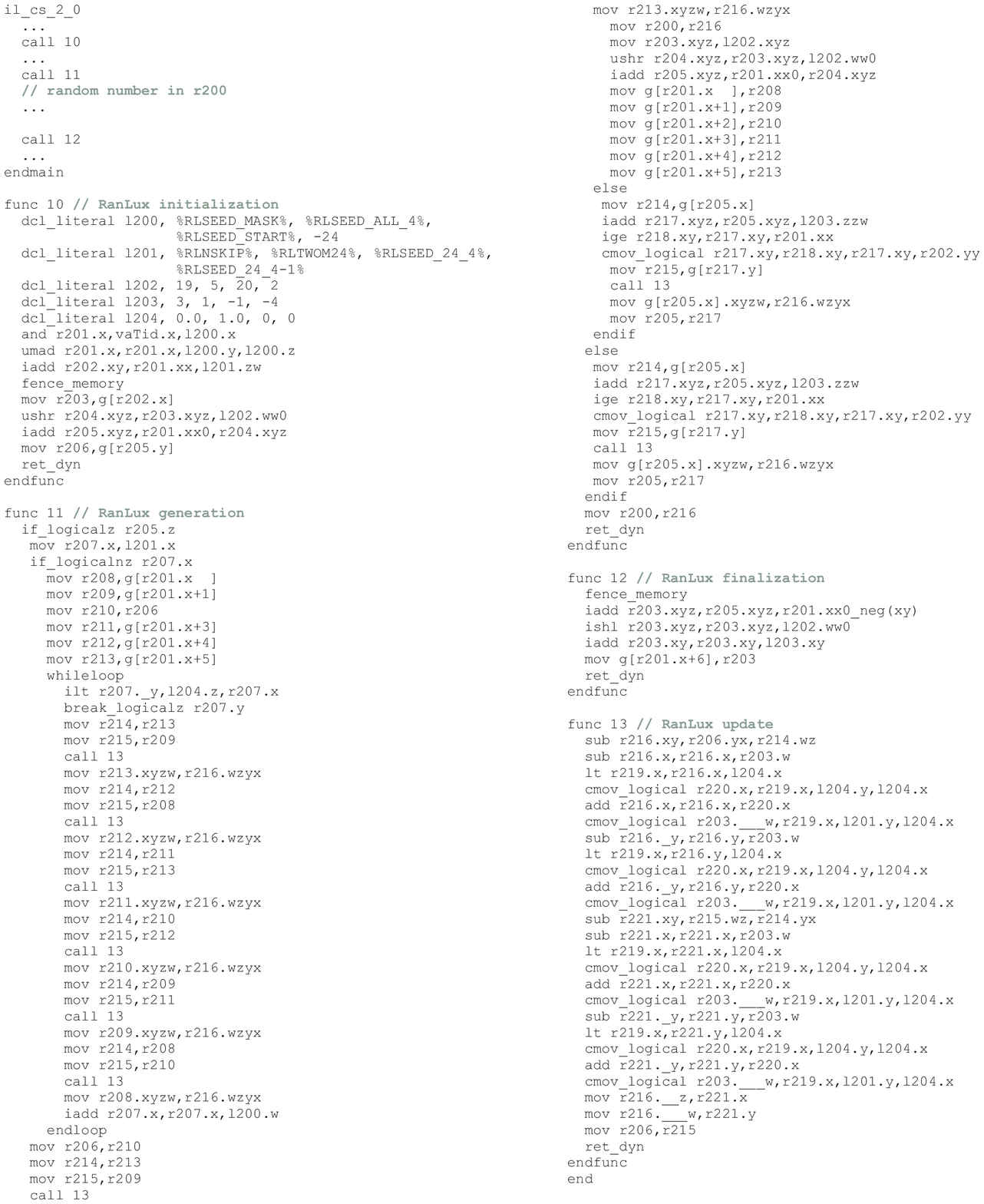}}
\caption{Source code of \PRNG{RANLUX} PRNG} \label{fig:RANLUX}
\end{center}
\end{figure}

To implement the \PRNG{RANLUX} generator, we use three quite
different approaches. First of all, the direct translation of
algorithm is performed. In this scheme all the seed values are
updated directly in GPU global memory. Every thread in point of
fact generates simultaneously the four independent sequences of
the pseudo-random numbers. Luxury is performed for all four
subthreads at the same time. This approach makes the algorithm
considerably simpler, but does not allow getting the best
performance.

Next evident approach is to use the indexed temporary array for
luxury operation. It allows to make process execution much faster:
at luxury level=3  3.5 times faster and 5 times faster at luxury
level=4, keeping the complexity of the algorithm meanwhile.

Third approach which really allows to make algorithm faster and in
practice minimize the dependence of the execution time on luxury
level, is the ``planar'' scheme. The geometry of the lag table is
taken into account as much as possible in this scheme. For
\PRNG{RANLUX} it consists of the 24 elements which naturally could
be grouped into six 4-component 32-bit registers. The new
difficulty to vectorize the \PRNG{RANLUX} algorithm is arisen due
to the necessity of the recursive calculation of the carry bit
$c_n$ which depends on the preceding states of the generator.
Therefore, some serial operations are appeared in planar
\PRNG{RANLUX} code as well as in presented here \PRNG{XOR128}
implementation. Planar \PRNG{RANLUX} procedure produces four
pseudo-random numbers from one sequence for one pass of the
generator.

Base algorithm \PRNG{RANLUX} requires discarding 24, 73, 199 and
365 values after one \PRNG{RANLUX} cycle for luxury level 1, 2, 3
and 4, respectively. However to perform luxury in the planar
implementation of the \PRNG{RANLUX}, it is convenient to discard
some larger number than ones, proposed by L\"uscher
\cite{Luscher:1993dy}. Strictly speaking, it is convenient to
discard the number of the values which are multiply by 24. In CPU
implementation such approach seems to be redundant (see
\cite{Luscher:1993dy}), but on GPU it shows better performance.
Thus, in planar scheme the following numbers are discarded: 24,
96, 216 and 384 (for luxury levels 1, 2, 3 and 4, respectively).

The ATI IL source code of planar \PRNG{RANLUX} PRNG is presented
in Figure \ref{fig:RANLUX}. The following variables are used
\begin{eqnarray}\nonumber
 \reg{RLSEED\_ALL\_4}=7, && \reg{RLTWOM24}=2^{-24},\\\nonumber
 \reg{RLSEED\_24\_4}=24/4=6, && \reg{RLSEED\_24\_4-1}=5,
\end{eqnarray}
\reg{RLSEED\_START} is the offset of the lag table in global
buffer, \reg{RLNSKIP} is the number of generated values to be
discarded (is defined by the luxury level), \reg{RLSEED\_\-MASK}
specifies the PRNG instance.

Due to relatively large lag table, \PRNG{RANLUX} as well as
\PRNG{RANMAR} generator requires the stand-alone initializing
procedure. This procedure is running only once and does not take
much resources, so its listing is not presented here. Also, it may
be realized in the host program with further copying the result to
the GPU global buffer.

Generator \PRNG{RANLUX}, as well as \PRNG{RANMAR} PRNG, possesses
the important feature in the context of the GPU realization -- the
generator kernel is build on floating-point arithmetic. Each
instance of the planar version of the \PRNG{RANLUX} requires
storing seven 4-component cells only (three indices which are
connected to each other, 24 items of lag table and carry bit). So
for 4096-thread run the size of the lag table is 448kB (or 1664kB
for the case of the global buffer and temporary indexed array
\PRNG{RANLUX} versions).

\subsection{Mersenne Twister}
The last PRNG which is implemented in this work is \PRNG{MT19937},
one of the Mersenne Twister generators family. This generator
seems to be very attractive nowadays due to its extremely long
period and relatively easy algorithm.

Mersenne Twister is incorporated in nVidia SDK as sample. The
realized in nVidia SDK version of the Mersenne Twister contains 19
element lag table and period about $2^{607}$. It is easy to
implement the \PRNG{MT19937} generator on the basis of this
example by substituting the relevant parameters.

In the presented realization of \PRNG{MT19937} we use the same
scheme as in the planar implementation of \PRNG{RANLUX}. Whole
624-element lag table is located into 156 four-component cells.
So, for the one pass of the PRNG it is easy to obtain four
sequential pseudo-random numbers. But unfortunately, it is
impossible to store whole lag table in the general-purpose
registers to perform the lag table updating procedure, because the
total number of the general-purpose registers allocated for one
thread is 128. Thus, all operations with the lag table are
realized with the slow direct access to the global buffer, what
greatly reduces the total performance of the generator.
Nevertheless, the using of the four-component elements somewhat
compensates this disadvantage.

\begin{figure}
\begin{center}
\resizebox{1.0\textwidth}{!}{\includegraphics{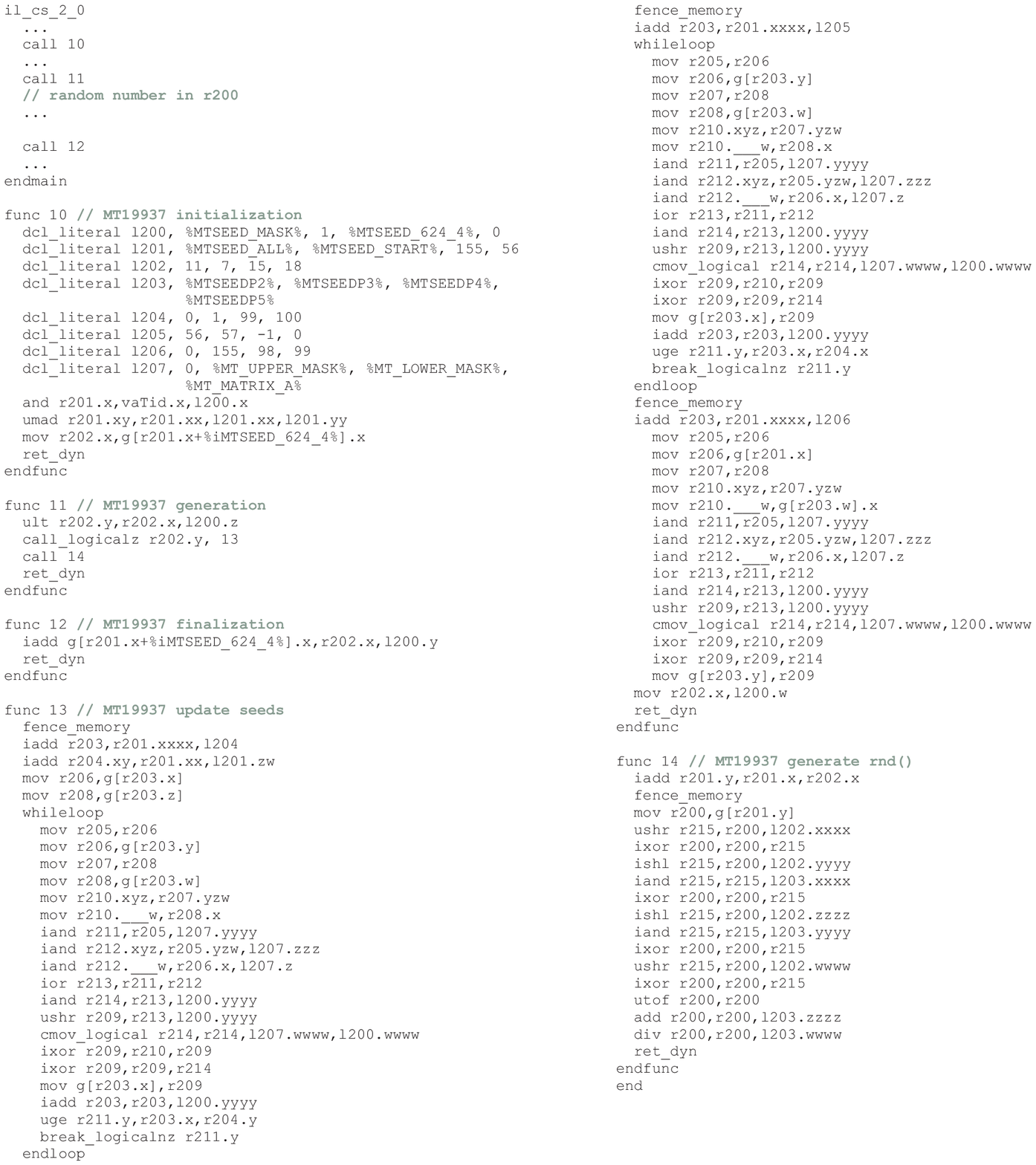}}
\caption{Source code of \PRNG{MT19937} PRNG} \label{fig:MT19937}
\end{center}
\end{figure}

The ATI IL source code of \PRNG{MT19937} PRNG is presented in
Figure \ref{fig:MT19937}. The following variables are used
\begin{eqnarray}\nonumber
 \reg{MTSEED\_624\_4}=156, && \reg{MTSEED\_ALL}=157,\\\nonumber
 \reg{MTSEEDP2}={\rm 9D2C5680_{16}}, && \reg{MTSEEDP3}={\rm EFC60000_{16}},\\\nonumber
 \reg{MTSEEDP4}=0.5, && \reg{MTSEEDP5}=4294967296.0,\\\nonumber
 \reg{MT\_UPPER\_MASK}={\rm 80000000_{16}}, && \reg{MT\_LOWER\_MASK}={\rm 7FFFFFFF_{16}},\\\nonumber
 \reg{MT\_MATRIX\_A}={\rm 9908B0DF_{16}},
\end{eqnarray}
\reg{MTSEED\_START} is the offset of the lag table in global
buffer, \reg{MTSEED\_MASK} specifies the PRNG instance.

The lag table of the \PRNG{MT19937} is the biggest one among the
lag tables of the presented generators. For 4096-thread run its
size is about 10MB. \PRNG{MT19937} also requires the
initialization procedure which prepares the lag table for the
first run.

\section{Performance results and discussion}\label{resultsdiscuss}
For all PRNGs implemented here we use the MS Visual Studio 2008
Express edition (C++ compiler) \cite{MSexp}. Original codes are
presented in corresponding literature (see Section
\ref{PRNGtheory}) and in several cases they are translated into
C++.

The CPU implementation of the PRNGs is constructed on the
following common scheme: each PRNG is implemented as subroutine
which produces only one pseudo-random number per call. Then the
main program sums up all produced values and checks the elapsed
time. The $10^8$ pseudo-random numbers are used for this
procedure. Elapsed time is averaged over several single thread
executions and the mean CPU performance is found. For the
reference PCs we use two machines:
\begin{itemize}
    \item Intel Core 2 Quad CPU Q6600 @ 2.40GHz {\small (L1 cache $4\times 32$KB,
    L2 cache $2\times 4$MB)},\\
    4GB RAM DDR2 400MHz {\small Dual Symmetric (5-5-5-16) Command rate 2T};
    \item Intel Celeron CPU 420 @ 1.60GHz {\small (L1 cache $32$KB, L2 cache
    $512$KB)},\\
    1.5GB RAM DDR2 333MHz {\small Dual (5-5-5-15) Command rate
    1T},
\end{itemize}
which correspond to the middle-level and entry-level computers,
respectively.

\begin{figure}
\begin{center}
\resizebox{0.9\textwidth}{!}{\includegraphics{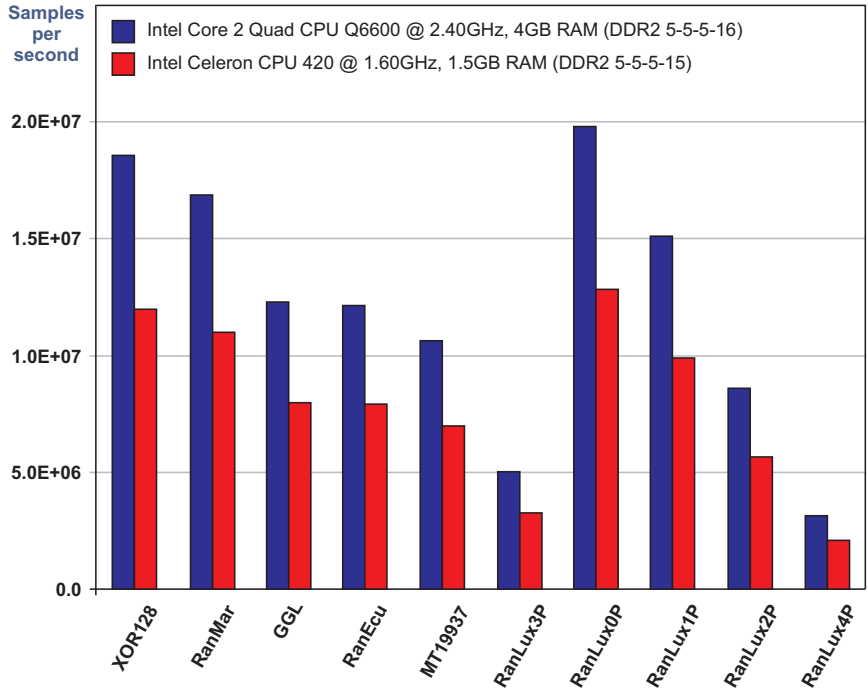}}
\caption{Performance results for some PRNGs on two PCs}
\label{fig:prng-cpu}
\end{center}
\end{figure}

The CPU performance results are presented in Figure
\ref{fig:prng-cpu} and Table \ref{PRNGGPUperformance}. It could be
seen that the differences among the performances of the generators
are about 5-6 times. \PRNG{XOR128} and \PRNG{RANMAR} show the best
performance results. Under CPU performance here and after we mean
the performance of the one-thread instance of the algorithm. Of
course, one-thread run does not provide maximal system
utilization, however it allows comparing the potential
performances of the systems. To show impartial assessment we can
just multiply CPU performance by the quantity of the threads,
supported up by peculiar processor, because it is always possible
to run several PRNG instances to produce different independent
pseudo-random sequences.

All the PRNGs implementations on GPU are carried out in ATI
Intermediate Language \cite{ATIIL} (ATI Catalyst 10.1 display
driver is used \cite{ATIdrv}). The host environment is also
realized in MS Visual Studio 2008 C++ \cite{MSexp}. All used here
GPUs are the main GPU devices installed in the system (i.e. they
also provide visualization for the operational system) which
lowers down the maximal performance of the system, but reflects
more precisely the usual configuration of the GPU computational
system. To obtain the performance of each PRNG, we run them up to
$1000$ times each to produce $4\times 10^7$ pseudo-random samples
on every pass. Each produced pseudo-random number is stored in
global buffer. Elapsed time is averaged over several executions.
The time spending to copy the initial input seeds to GPU memory
and final mapping the GPU memory into host memory is not taken
into account.

\begin{table}
\caption{The performance results of the presented here PRNG
implementations on different ATI GPUs (here {\bf CPU} is Intel
Core 2 Quad CPU Q6600 at 2.40GHz and {\bf CPU$_2$} is Intel
Celeron CPU 420 at 1.60GHz)}\label{PRNGGPUperformance}\small
\begin{center}
\begin{tabular}{|l|c|c|c||c|c||c|}\hline
 \rule{0pt}{10pt} & \multicolumn{3}{c||} {$\times 10^9$ per second} & \multicolumn{2}{c||} {$\times
 10^7$ per second} & {\bf Speed-up}\\\cline{2-6}
 \rule{0pt}{10pt} {\bf PRNG} & {\bf 5850} & {\bf 4870} & {\bf 4850} & {\bf CPU} & {\bf CPU$_2$} &
 {\bf factor}\\\hline
 \rule{0pt}{10pt} \PRNG{GGL}
 & $8.37$ & $5.05$ & $4.21$ & $1.23$ & $0.80$ & 681\\\hline
 \rule{0pt}{10pt} \PRNG{XOR128}
 & $8.45$ & $6.29$ & $4.52$ & $1.86$ & $1.20$ & 455\\\hline
 \rule{0pt}{10pt} \PRNG{RANECU}
 & $4.98$ & $3.32$ & $2.66$ & $1.21$ & $0.79$ & 411\\\hline
 \rule{0pt}{10pt} \PRNG{RANLUX3P}
 & $1.08$ & $1.02$ & $0.63$ & $0.50$ & $0.33$ & 216\\\hline
 \rule{0pt}{10pt} \PRNG{RANLUX4P}
 & $1.02$ & $0.86$ & $0.58$ & $0.32$ & $0.21$ & 322\\\hline
 \rule{0pt}{10pt} \PRNG{MT19937}
 & $0.50$ & $0.62$ & $0.36$ & $1.07$ & $0.69$ & 47\\\hline
 \rule{0pt}{10pt} \PRNG{RANMAR}
 & $0.18$ & $0.23$ & $0.14$ & $1.69$ & $1.10$ & 11\\\hline
\end{tabular}
\end{center}
\end{table}

\begin{figure}
\begin{center}
\resizebox{0.9\textwidth}{!}{\includegraphics{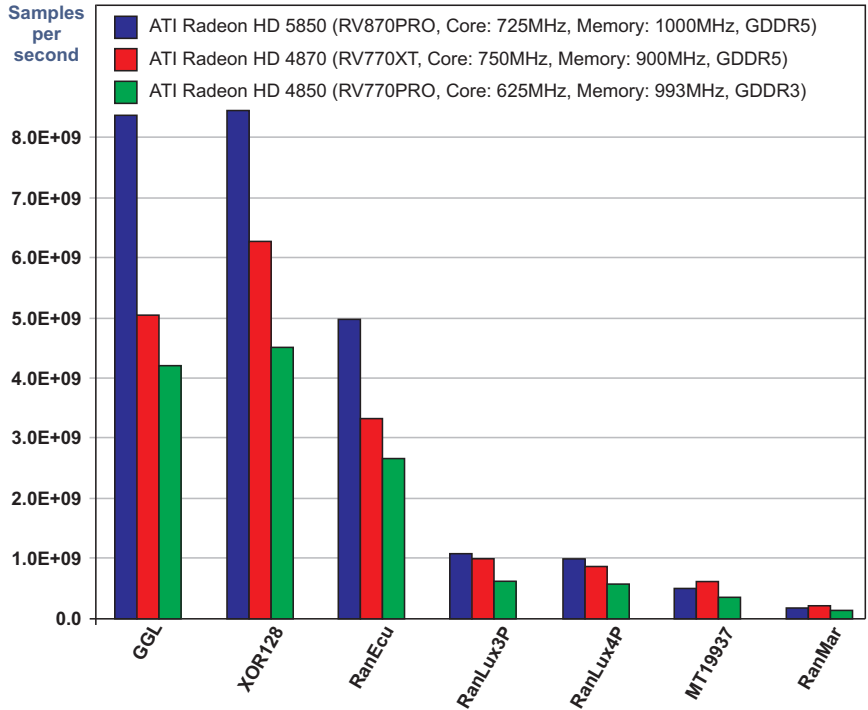}}
\caption{Performance results for some PRNGs on different GPUs}
\label{fig:prng-gpu}
\end{center}
\end{figure}

The performance results are collected in Figure \ref{fig:prng-gpu}
and Table \ref{PRNGGPUperformance}. The first column of the Table
\ref{PRNGGPUperformance} contains the name of the implemented
generator. The number pseudo-random numbers which could be
produced by corresponding PRNG per one second on corresponding GPU
are show in the columns 2-4. Here {\textsf{HD5850}},
{\textsf{HD4870}} and {\textsf{HD4850}} are the ATI Radeon video
cards. The next two columns contain the same information obtained
on two mentioned CPUs. The last column shows the speed-up factor
of the using the ATI Radeon {\textsf{HD5850}} in compare with the
using of the Intel Core 2 Quad CPU Q6600 at 2.40GHz.

Memory access is a bottleneck of the GPU-applications. It is
confirmed once again by different PRNG performance results on
different ATI GPU hardware. The PRNGs with the greater number of
the memory operations demonstrate the worst performance results.

In the present realizations \PRNG{GGL} and \PRNG{XOR128}
generators require to keep only one 4-component seed-value per
thread. This directly influences their work -- both generators
showed the best productivity. Despite the generator \PRNG{XOR128}
in the present code has sequential part which slightly lowers the
speed of the generator operation its performance turned out to be
the highest. It may be explained by the fact that only bit
operations are used in the arithmetic kernel of the \PRNG{XOR128}
generator which along with floating-point operations are the
fastest implemented on ATI hardware. Generator \PRNG{GGL} is
realized on the base of the integer scheme of Park and Miller
\cite{Park:1988rn} which slightly brings down its output in
compare with \PRNG{XOR128}, in spite of less quantity of the
intermediate operations.

The \PRNG{RANECU} generator already requires keeping two
seed-values for its operation. Along with using integer
arithmetical operations in \PRNG{RANECU} kernel it somewhat lowers
its performance in compare with \PRNG{GGL} and \PRNG{XOR128}
generators. The algorithm simplicity, the performance up to
$5\times 10^9$ samples per second and the considerably better
statistical properties allow extensively using \PRNG{RANECU}
generator on GPU. In order to increase the \PRNG{RANECU} period it
is reasonable to employ the MRG based on the combinations not two
but three MLCGs. However this extension requires the additional
study of the new generator statistical properties. It may be a
subject of the further research in this field.

The \PRNG{RANLUX} generator shows considerably high performance on
GPU. To a significant degree it may be explained by fortunate
matching of GPU architecture and the generator parameters. It is
managed to minimize generator dependence on decorrelating luxury
procedure by the actual organization of the virtual cache. Thereby
it is possible to use maximal luxury level with the minimal
performance penalty in practical applications. In the opposition
to CPU realization where the difference between the performances
at luxury level=0 and luxury level=4 reaches 6 and more times, on
GPU this parameter is only 10-23\% (for different hardware).

It is obviously that performance advantage also depends on the way
of algorithm realization. In order to demonstrate this fact, we
showed in Table \ref{RANLUXperformance} the performance values of
different realization of the same algorithm for \PRNG{RANLUX}
generator for a number of GPUs. \PRNG{RANLUX} implementation
through the global buffer is the slowest one. In fact the present
realization of algorithm entirely reproduces the dependence of the
GPU realization on luxury level. Acceleration in compare with CPU
is achieved only because of such GPU parameters as number of
stream cores and engine clock rate. An unexpected result was the
fact that realization of for global buffer realization of
\PRNG{RANLUX} {\textsf{HD4870}} GPU shows better performance than
{\textsf{HD5850}}. It is closely related to the fact that memory
data access in the kernel is organized not in the best way and a
lot of time is spent to synchronize memory access operations.
Obtaining better performance is possible by lowering the
instructions density which refer to the global buffer.

\begin{table}
\caption{Performance results of the different implementations of
\PRNG{RANLUX} generator on some ATI GPUs (ATI Radeon
{\textsf{HD5850}}, {\textsf{HD4870}}, {\textsf{HD4850}}), $\times
10^8$ pseudo random numbers per
second}\label{RANLUXperformance}\small
\begin{center}
\begin{tabular}{|c|c|c|c||c|c|c||c|c|c|}\hline
 \rule{0pt}{10pt} luxury & \multicolumn{3}{c||} {Planar} & \multicolumn{3}{c||}
 {Temporary array} & \multicolumn{3}{c|} {Global buffer}\\\cline{2-10}
 \rule{0pt}{10pt} level & {\bf 5850} & {\bf 4870} & {\bf 4850} & {\bf 5850} &
 {\bf 4870} & {\bf 4850} & {\bf 5850} & {\bf 4870} & {\bf 4850}\\\hline
 \rule{0pt}{10pt} 0 & 11,35 & 10,64 & 6,54 & 4,07 & 5,56 & 3,35 & 5,02 & 6,92 &
 4,32\\\hline
 \rule{0pt}{10pt} 1 & 11,00 & 10,14 & 6,19 & 4,05 & 5,39 & 3,33 & 3,50 & 4,66 &
 2,71\\\hline
 \rule{0pt}{10pt} 2 & 11,01 & 10,17 & 6,20 & 4,02 & 5,00 & 3,23 & 2,18 & 2,61 &
 1,55\\\hline
 \rule{0pt}{10pt} 3 & 10,85 & 10,17 & 6,27 & 3,85 & 4,18 & 2,83 & 1,11 & 1,20 &
 0,72\\\hline
 \rule{0pt}{10pt} 4 & 10,24 & 8,62 & 5,77 & 3,51 & 3,44 & 2,42 & 0,68 & 0,71 &
 0,43\\\hline
\end{tabular}
\end{center}
\end{table}

Second \PRNG{RANLUX} realization, through the indexed temporary
array, is a bit faster than the first one. But the same situation
with {\textsf{HD4870}} and {\textsf{HD5850}} GPUs performances can
be observed here as well. The present realization has sense only
for luxury levels 2-4, when time share, which is spent on indexed
temporary array preparation, is compensated by luxury operation
saving time itself. The main strong feature of this implementation
is the complete correspondence to the classic algorithm, published
in \cite{James:1993vv}.

The last presented implementation of \PRNG{RANLUX} (which is
called planar here) possesses the best performance due to maximal
reduction of memory access operations quantity. General advantage
of performance on GPU in compare with CPU is up to 322 times
(highest luxury level = 4). The modified algorithm is used here
(see section \ref{RANLUXimplementation}) at which a bit higher
quantity of pseudo-random numbers after one PRNG cycle are
discarded away for decorrelation of lag table elements, than it is
set up in the classic algorithm.

In the first two \PRNG{RANLUX} realizations of the algorithm, it
requires to read four memory cells (not accounting for the luxury
procedure operations) and write three cells (two lag table items,
carry bit and indices) to obtain one pseudo-random number (updated
lag table item, new carry bit and new indices). In planar scheme
the carry bit occupies one of the four-component cells with the
indices, so it needs less quantity of reading and writing
operations (four read and two writes).

The performance of the shown code of the \PRNG{MT19937} generator
is turned out to be rather low. The main factor is the size of the
lag table, the update of which is required after each PRNG cycle.
Planar scheme applied here allows increasing the generator
performance roughly in four times in compare with the direct
four-threads realization, but the quantity of the memory access
operations remains rather high which does not allow to achieve
acceptable results. It is possible to reduce the lag table size by
choosing another generator from the Mersenne Twister family. But
from one hand, \PRNG{MT19937} generator parameters allow to
realize the planar scheme (proved to be a good here), and from
another hand there is a task to build the analogue of the actually
used generators on GPU while algorithms realization. Although to
generate one pseudo-random number it is needed only two read
operations and one write only, the lag table update procedure
drastically decelerates the generator.

The worst performance is shown by \PRNG{RANMAR} generator which
differs by the large enough lag table and high memory access
operation density (four reads and three writes).

All presented generators realizations allow to produce at slight
changes pseudo-random numbers with double precision either by
using the pairs of the generated numbers or directly through the
double precision conversion. The last method is not fully correct,
because it considerable lowers the quantity of the possible
realization of the obtained double precision number.

Paying attention to the \PRNG{XOR128} performance one can asserts
that L'Ecuyer generator \PRNG{Seven-XORShift}
\cite{Lecyuer:2005xs} is seemed to be very promising for
realization on GPU due to the similarity to the \PRNG{XOR128}
structure except for the eight-element lag table.

The periods of the \PRNG{XOR128}, \PRNG{RANLUX}, \PRNG{RANMAR} and
\PRNG{MT19937} generators are considered to be unachievable in
medium-term perspective even for GPU clusters. Meanwhile the
period of the \PRNG{GGL} generator can be exhausted in a split
second even on one relatively old GPU. So, it is necessary to pay
special attention to the generator applicability borders while
developing applications on GPU.

Undoubtedly the presented algorithms realization on the platform
independent level such as OpenCL will be very important, but it
goes beyond the present work frame and is a task for future
research. The obtained results of the generators performances make
it interesting to use the GPU for the investigation of the
generators statistical properties.

\section{Conclusions}\label{conclusions}
In the present paper the most popular uniform pseudo-number
generators which are used in Monte Carlo simulations in
high-energy physics are realized on GPU. The list of the modern
software packages with the indication of the generators used is
represented. A theoretical background for pseudo-random number
generation is described. The source codes of the implemented
generators (multiplicative linear congruential (\PRNG{GGL}),
XOR-shift (\PRNG{XOR128}), \PRNG{RANECU}, \PRNG{RANMAR},
\PRNG{RANLUX} and Mersenne Twister (\PRNG{MT19937})) are
shown\footnote{ When the present paper was prepared for
publication, a new version of ATI Stream SDK v.2.01 has been
released which includes the Park-Miller generator with Bays-Durham
shuffle and the Mersenne Twister generator}.

The comparative analysis of the PRNG performances on CPU and ATI
GPU is presented. The obtained speed-up factor is hundreds of
times higher as compare to CPU. Performance analysis of the
mentioned generators with taking into account their statistical
properties allows to conclude that the most appropriative
generator for Monte Carlo simulations on GPU is planar
\PRNG{RANLUX} with luxury level 4. Offered planar scheme makes it
possible to increase significantly the performance of the most
generators by the reducing memory access operations.

Offered PRNG program model (generator division into separate
subroutines) also allows to obtain additional considerable gain of
performance at the multiple calls of the PRNG generating
subprogram. This is achieved by means of a substantial reduction
in the number of the memory access operations per PRNG cycle.

In the present paper the generator performances in different
realizations are analyzed by way of \PRNG{RANLUX} PRNG example. It
is shown once again that the memory access operations are the
bottleneck of GPUs. It is possible to conclude that for GPU
implementations of PRNGs it is better to choose algorithms with
the minimal lag table size (which to be advisable multiply by 4)
perhaps in spite of some complication of algorithms.

Potential of GPU implementation in Monte Carlo simulations is not
fully realized nowadays. This is despite of large amount of works
dedicated to this problem. GPU sector development dynamics makes
it possible to predicate that this trend is one of the most
promising nowadays.

\section*{Acknowledgments}
Author thanks to Alexander Lukashenko for the inspiration of
GPGPU-technology, Michael Bordag and Wolfhard Janke for the
organization of the mini-workshop {\it ``Simulations on GPU''},
where the subject of pseudo-random numbers generation on GPUs was
discussed. We would also like to thank Vladimir Skalozub and Eugen
Setov for essential help with the paper preparation.

\appendix
\section{Basic classes of PRNGs}\label{PRNGtheory}
In this section we briefly provide the theoretical background for
the main classes of the pseudo-random number generators. There are
numerous books and reviews about the subject, so the details could
be found in the references.

While pseudo-random numbers generation, there are only two
internal sources of the randomness which could be used in the
algorithm. They are the sequence itself (previous values in a
sequence precisely) and the starting parameters (PRNG parameters
and seed values). Actually, the algorithms of the PRNGs are
distinguished by the way of the using these sources. So,
particular PRNG is a certain function $f$ which produces the next
value $X_n$,
\begin{eqnarray}
X_{n}^{\PRNG{PRNG}} = f(X_{n-1},X_{n-2},\cdots,X_{n-r}).
\end{eqnarray}
Here and below we will use upper index \PRNG{PRNG} to identify the
sequence produced by particular generator \PRNG{PRNG}. The maximum
period of the generator is the length of the cyclic sequences
produced by PRNG and is limited by the number of the states that
can be represented by PRNG.

First simple PRNGs uses only one previous value $X_{n-1}$ ($r=1$)
for generation, but in such scheme the PRNG period is limited by
the bit capacity of the machine. If one uses the longer tables of
the sequence of the previous values, from one hand it makes the
generator period longer and its statistics properties better, as
well as allows to simplify transformation function $f$ for getting
better output of the generator. From the other hand, longer tables
require more complicated generator initialization and reduce its
portability (the strict control of the architecture is needed, for
example, the size of the cache memory), and it is obviously
necessary to have much more memory to store such tables. A good
generator is always a golden middle between algorithm complexity,
the statistical properties of the generator and the size of the
seed table.

According to the basic requirement for PRNG -- repeatability, any
starting state of the generator may cause only one specified
sequence. The initialization of the seed table is often made by
simpler generator which has lower requirements. Most of all linear
congruential generators or their combination is used for
initialization. So, it allows to set starting generator states by
the limited set of the input seed values (often 1-2 numbers). But
Marsaglia \cite{Marsaglia:2003rn} drew our attention to the fact
that the quantity of the starting states decreases drastically.
For example, Mersenne Twister generator \PRNG{MT19937} uses
624-element seed table (which may contain about
$(2^{32})^{624}\simeq 10^{6011}$ different values) whenever 32-bit
number is usually used for its initialization, for which only
$(2^{32})\simeq 4\times 10^9$ possible values are given.
Certainly, algorithm contains the possibility of the vector
initialization. Nevertheless, only one seed is used to simplify
the initialization.

The period of PRNG nearly always depends on its parameters and
seed values. Thus, we always show only the upper bound for the
value of the generator period.

\subsection{Linear congruential generators}\label{PRNGLCG}
Linear congruential generators ({\it LCG}) is one of the oldest
and most popular class of the PRNGs, which is widely used in
computations in particular due to the encyclopedic work of Knuth
\cite{Knuth:1997ap}. It is based on so-called linear congruential
integer recursion,
\begin{eqnarray}\label{LCG}
X_n^{\PRNG{LCG}} = (a X_{n-1} + c)\mod m,
\end{eqnarray}
where increment $c$ and modulus $m$ are desired to be positive
coprime integers ($c<m$) to provide a maximum period, multiplier
$a$ is an integer in the range $[2;(m-1)]$. If increment $c=0$ the
LCG is often called the multiplicative linear congruential
generator ({\it MLCG}),
\begin{eqnarray}\label{MLCG}
X_n^{\PRNG{MLCG}} = a X_{n-1}\mod m.
\end{eqnarray}

The maximum period of LCG $P_{\PRNG{LCG}}$ strongly depends on LCG
parameters and
\begin{eqnarray}\label{LCGPeriod}
P_{\PRNG{LCG}}\leq (m-1).
\end{eqnarray}
Demonstrative situation with poor choice of LCG parameters
happened with infamous generator \PRNG{RANDU} - MLCG($a=2^{16}+3$,
$m=2^{31}$),
\begin{eqnarray}\label{RANDU}
X_n^{\PRNG{RANDU}}= 65539 X_{n-1}\mod 2147483648,
\end{eqnarray}
which suffers from three-point correlations among the sequential
elements:
\begin{eqnarray}
X_n^{\PRNG{RANDU}}=6X_{n-1}-9X_{n-2}.
\end{eqnarray}

Another well-known LCG is the standard 48-bit generator
\PRNG{DRAND48}, which is a LCG($a=25214903917$, $c=11$,
$m=2^{48}$)
\begin{eqnarray}\label{DRAND48}
X_n^{\PRNG{DRAND48}} = \left(25214903917 X_{n-1} + 11\right)\mod
2^{48}.
\end{eqnarray}

Park and Miller propose \cite{Park:1988rn} a portable minimal
standard Lehmer generator\footnote{The original parameters of
Lehmer generator are $a=23$, $m=10^8+1$} \cite{Lehmer:1951mm}
known as prime modulus ~multiplicative ~linear ~congruential
~generator (\PRNG{PMMLCG}). Sometimes it is also denoted by the
acronyms \PRNG{RAN0}, \PRNG{CONG}, \PRNG{SURAND} or \PRNG{GGL}.
Park and Miller choose the Mersenne prime number $2^{31}-1$ as the
modulus $m$, multiplier $a=7^5$ and increment $c=0$,
\begin{eqnarray}
X_n^{\PRNG{GGL}} = 16807 X_{n-1}\mod 2147483647.
\end{eqnarray}
The total period of the \PRNG{GGL} is relatively short,
$P_{\PRNG{GGL}}=(2^{31}-1)-1=2147483646$.

One of the main well-known problems of the LCG is that every LCG
produces $n$-tuples of uniform variates which lie in at most
parallel hyperplanes \cite{Marsaglia:1968rf,Marsaglia:1972sl}. The
other defects of the LCG are:
\begin{itemize}
    \item the dependence of the generator period on initial seed;
    \item the influence of the chosen modulus on statistical
    properties of the pseudo-random sequence;
    \item lowest bits are not random.
\end{itemize}
And finally, the minimal integer value produced by MLCG is 1, not
0.

\subsection{Feedback shift register generators}\label{PRNGFSRG}
In 1965 Tausworthe \cite{Tausworthe:1965fs} introduced a new
generator based on bit sequence
\begin{eqnarray}\label{LFSRseq}
X_{n}^{\PRNG{LFSR}}=\left(\sum_{i=1}^{r}a_iX_{n-i}\right) \mod
2,~~
\end{eqnarray}
where $a_i=\{0,1\}$, $X_i=\{0,1\}$, $r\leq n$, is called linear
feedback shift register algorithm ({\it LFSR}).

The period of the LFSR is the smallest positive integer $P$ for
which
\begin{eqnarray}
\left(X_{r-1},\cdots,X_0\right)=\left(X_{P+r-1},\cdots,X_P\right).
\end{eqnarray}

The next state could be obtained with the following transformation
\begin{eqnarray}
\left(
\begin{array}{c}
X_{n}\\
X_{n-1}\\
\vdots\\
X_{n-r+1}
\end{array}
\right)=\left(
\begin{array}{cccc}
a_1 & \cdots & a_{r-1} & a_r\\
1 & \cdots & 0 & 0 \\
\vdots & \ddots & \vdots & \vdots\\
0 & \cdots & 1 & 0
\end{array}\right)
\left(
\begin{array}{c}
X_{n-1}\\
X_{n-2}\\
\vdots\\
X_{n-r}
\end{array}
\right)
\end{eqnarray}
or equivalently for initial state it is
\begin{eqnarray}
\left(
\begin{array}{c}
X_{n}\\
X_{n-1}\\
\vdots\\
X_{n-r+1}
\end{array}
\right)=\left(
\begin{array}{cccc}
a_1 & \cdots & a_{r-1} & a_r\\
1 & \cdots & 0 & 0 \\
\vdots & \ddots & \vdots & \vdots\\
0 & \cdots & 1 & 0
\end{array}\right)^{n-r+1}
\left(
\begin{array}{c}
X_{r-1}\\
X_{r-2}\\
\vdots\\
X_{0}
\end{array}
\right).
\end{eqnarray}

The characteristic polynomial of $r\times r$ transformation matrix
$A$
\begin{eqnarray}
A=\left(
\begin{array}{cccc}
a_1 & \cdots & a_{r-1} & a_r\\
1 & \cdots & 0 & 0 \\
\vdots & \ddots & \vdots & \vdots\\
0 & \cdots & 1 & 0
\end{array}\right)
\end{eqnarray}
is
\begin{eqnarray}
a(x)=\determ\left(A-xI\right)=x^r-\sum_{i=1}^{r} a_i
x^{r-i},~~a_r=1,
\end{eqnarray}
where $I$ is an identity matrix.

According to the theory of the finite fields \cite{Lidl:1986ff}
the maximal period $P_{\PRNG{FSRG}}=2^r-1$ is achieved if and only
if the characteristic polynomial $a(x)$ is an irreducible
polynomial over Galois field $GF(2)$. In other words, if and only
if the smallest positive integer $p$: $(x^p \mod a(x)) \mod 2=1$
is $p=2^r-1$ \cite{Lecuyer:2004hc}.

For computational efficiency, most of the $a_i$ in (\ref{LFSRseq})
should be zero. In $GF(2)$ there is only one irreducible binomial,
$x+1$, which would yield an unacceptable period
\cite{Gentle:2003}. Consequently, the trinomials are usually used
to express the recurrence sequence (\ref{LFSRseq}),
\begin{eqnarray}\label{FSRtrin}
X_n=\left(X_{n-r}+X_{n-s}\right) \mod 2,~~~s<r<n.
\end{eqnarray}
Addition modulo 2 for one-bit variables is ordinary binary
exclusive-or operation XOR, so (\ref{FSRtrin}) may be rewritten as
\begin{eqnarray}\label{FSRXOR}
X_n=X_{n-r}\XOR X_{n-s}.
\end{eqnarray}

LFSR recursion (\ref{LFSRseq}) produces pseudo-random bit
sequence. To obtain $k$-bit pseudo-random integers $Y_n$ from such
recursion, one can group up $k$ sequential bits,
\begin{eqnarray}
Y_n=\sum_{j=1}^{k}2^{k-j}X_{kn+j-1}.
\end{eqnarray}
Such method is called the digital multistep method of Tausworthe
\cite{Niederreiter:1992qm}.

Another method proposed by Lewis and Payne \cite{Lewis:1973gs} is
the generalized feedback shift register ({\it GFSR}). In GFSR
scheme bits in the positions $j$ of the pseudo-random integer are
filled with the copy of initial one-bit recursion (\ref{FSRXOR})
which has a period $2^r-1$ with some nonnegative offsets $d_j$,
\begin{eqnarray}
Y_n=\sum_{j=1}^{k}2^{k-j}X_{n+d_j}.
\end{eqnarray}
Clearly, LFSR is the particular case of GFSR.

For example of GFSR it could be mentioned the \PRNG{R250}
generator \cite{Kirkpatrick:1981sr}. It is another infamous PRNG
which causes the severe problems in the Monte-Carlo simulations of
the two-dimensional Ising model using the single-cluster Wolff
update algorithm (see \cite{Janke:2002} and references therein).
For \PRNG{R250} the GFSR parameters are $r=250$ and $s=103$,
\begin{eqnarray}
X_n^{\PRNG{R250}}=X_{n-250}\XOR X_{n-103}.
\end{eqnarray}
The \PRNG{R250} period is $P_{\PRNG{R250}}=2^{250}-1\approx
1.81\times 10^{75}$.

\subsection{Lagged Fibonacci generators}\label{PRNGLFG}
Lagged Fibonacci generators ({\it LFG}) is another class of the
PRNGs. It is based on the well-known Fibonacci recurrence sequence
\begin{eqnarray}\label{Fibonacci}
X_n=X_{n-1}+X_{n-2}.
\end{eqnarray}
Because the simple Fibonacci generator is not very good
\cite{James:1988vf} one always uses the generalized relation
(\ref{Fibonacci}) with respect to any given binary arithmetic
operation $\odot$ and prehistory
\begin{eqnarray}
X_n^{\PRNG{LFG}}=(X_{n-r}\odot X_{n-s}) ~{\rm mod}~m,
\end{eqnarray}
where $r$ and $s$ are called ``lags'', $r\leq n$ and $1<s<r$.

The LFG period $P_{\PRNG{LFG}}$ for different operations $\odot$
is
\begin{eqnarray}\label{LFGPeriods}
P_{\PRNG{LFG}}\leq \Biggl\{
\begin{array}{ll}
(2^r-1)m/2 & {\rm for~ +~ or~ -}\\
(2^r-1)m/8 & {\rm for~ \times}\\
(2^r-1) & {\rm for~ XOR}
\end{array}.
\end{eqnarray}

The main attractive features of the LFGs are long period,
potential absence of conversion integer into float operations and
simple recursive scheme which not requires the heavy mathematical
operations. However, for generation LFGs it is needed to store $r$
previous pseudo-random values.

There are numerous possible pairs of the LFG lags
\cite{Knuth:1997ap}. The larger lags lead to the decreasing of the
correlations between the numbers in the sequence. But even for
relatively short lag table $r\gtrsim 20$ it passes many
statistical tests.

\PRNG{RAN3} generator \cite{Knuth:1997ap} could be mentioned as an
example of the LFG. It was proposed by Mitchell and Moore
(unpublished), with lags $r=55$ and $s=24$, $m=10^9$ and operation
subtraction,
\begin{eqnarray}
X_n^{\PRNG{RAN3}}=\left(X_{n-55}-X_{n-24}\right)\mod 10^9.
\end{eqnarray}
The \PRNG{RAN3} period is $P_{\PRNG{RAN3}}=(2^{55}-1)
10^9/2\approx 1.8\times 10^{25}$.

\subsection{Combined generators}\label{PRNGCG}
Combined generators are a special class of the PRNGs, which
contains the features of the different PRNG classes. There are two
main motivations to use combined generators:
\begin{itemize}
    \item the period increasing of the generator,
    \item the improving of the generator statistical properties.
\end{itemize}

\subsubsection{Multiple recursive generators}\label{PRNGMRG}
The obvious extensions of the LCG is the multiple recursive
generator ({\it MRG}) \cite{Lecuyer:1988ep,Lecuyer:1990si} which
is determined as the combination of the MLCGs
\begin{eqnarray}
X_n^{\PRNG{MRG}}=\left(a_1X_{n-1}+a_2X_{n-2}+\ldots+a_kX_{n-k}+c\right)\mod
m.
\end{eqnarray}
When $k>1$, MRG is usually called MRG of the $k$ order. The
maximal period of the MRG is $P_{\PRNG{MRG}}\leq m^k - 1$. In fact
LFG is the special case of the MRG (for multipliers $a_i=1$ and
$c=0$).

By decomposition of the modulus of the MLCG into two terms
$m=aq+r$ and eqn.(\ref{MLCG}) may be written as following
\cite{Lecuyer:1988ep}
\begin{eqnarray}\label{RANECU1}
 X_n=aX_{n-1}\mod m=\left(a\left(X_{n-1}\mod q\right)-\lfloor X_{n-1}/q\rfloor
 r\right)\mod m,\\\nonumber
 X_n=X_n+m~~ {\rm for}~X_n<0,
\end{eqnarray}
where $\lfloor a/b \rfloor$ denotes the integer part of the
$(a/b)$ and
\begin{eqnarray}
 q=\lfloor m/a\rfloor,~~~ r=m\mod a.
\end{eqnarray}

To provide the uniform distribution the combinations of $l$
generators (\ref{RANECU1}) may be combined as
\cite{Lecuyer:1988ep}
\begin{eqnarray}
 X_n=\left(\sum_{j=1}^l (-1)^{j-1}X_{j,n}\right) \mod (m_1-1),\\\nonumber
 X_n=X_n+(m_1-1)~~ {\rm for}~X_n \leq 0.
\end{eqnarray}
Here the new index $j$ in $X_{j,n}$ means the $n$-th value
(\ref{RANECU1}) of $j$-th generator.

One of the possible MRGs is the \PRNG{RANECU} generator
\cite{Lecuyer:1988ep}. It is the combination of two MLCGs ($l=2$)
with $a_1=40014$, $m_1=2147483563$ and $a_2=40692$,
$m_2=2147483399$. The \PRNG{RANECU} period is
$P_{\PRNG{RANECU}}=(m_1-1)(m_2-1)/2\approx 2.30584\times 10^{18}$.
MRGs have good statistics and pass most the tests.

\subsubsection{XORShift}\label{PRNGXOR}
\PRNG{XORShift} PRNG, proposed by Marsaglia
\cite{Marsaglia:2003xr}, is another member of the GFSR generators
class. Let $X_0$ be a some initial $k$-bit row-state of
\PRNG{XORShift} and $T$ is $k\times k$ nonsingular binary matrix
which sets linear transformation. The $n$-th PRNG state may be
derived through the following equation
\begin{eqnarray}
X_n^{\PRNG{XORShift}}=X_0T^n.
\end{eqnarray}
To ensure the performance requirements Marsaglia proposed the
special form of matrix $T$,
\begin{eqnarray}
T=\left(I+L^a\right)\left(I+R^b\right)\left(I+L^c\right),
\end{eqnarray}
where matrices $L$ and $R$ are $k\times k$ binary matrices which
effect shift of one to the left and right, correspondingly. So, if
$X_m$ is a $k$-bit state then $L^a$ causes the new state
$L^aX_m\equiv (X_m\ll a)$ as well as $(I+L^a)$ -- the state
$(I+L^a)X_m\equiv X_m\XOR (X_m\ll a)$. In \cite{Marsaglia:2003xr}
Marsaglia lists all possible full-period triplets $(a,b,c)$ for
32-bit (648 combinations) and 64-bit (2200 combinations)
\PRNG{XORShift} PRNG.

The maximal period of \PRNG{XORShift} is
\begin{eqnarray}
P_{\PRNG{XORShift}}\leq 2^k-1.
\end{eqnarray}

In spite of \PRNG{XORShift} PRNG passes the \TEST{DIEHARD Battery
of Tests of Randomness} \cite{Marsaglia:2003xr} L'Ecuyer appoints
that it ``spectacular failed'' the \TEST{SmallCrush} and
\TEST{Crush} tests \cite{Lecyuer:2005xs}. L'Ecuyer does not
recommend to use this class of the generators, but proposes the
own version of the \PRNG{XORShift} implementation --
\PRNG{Seven-XORShift}.

Marsaglia gives an example of the \PRNG{XORShift} generator for
128-bit vector with four 32-bit components -- \PRNG{XOR128} PRNG
\cite{Marsaglia:2003xr},
\begin{eqnarray}
\left(X_{n-3},X_{n-2},X_{n-1},X_{n}\right)^\PRNG{XOR128}=
\left(X_{n-4},X_{n-3},X_{n-2},X_{n-1}\right)\cdot\\\nonumber \cdot
\left(
\begin{array}{cccc}
0 & 0 & 0 & (I+L^{11})(I+R^{8})\\
I & 0 & 0 & 0\\
0 & I & 0 & 0\\
0 & 0 & I & (I+R^{19})
\end{array}
\right),
\end{eqnarray}
or in the terms of the 32-bit components
\begin{eqnarray}
t=(X_{n-4}\XOR(X_{n-4}\ll 11)),\\\nonumber
X_{n-3}=X_{n-2},~~X_{n-2}=X_{n-1},~~X_{n-1}=X_{n},\\\nonumber
X_{n}=(X_{n-1}\XOR(X_{n-1}\gg 19))\XOR(t\XOR(t\gg 8)).
\end{eqnarray}
The \PRNG{XOR128} period is $P_{\PRNG{XOR128}}\leq 2^{128}-1$.

\subsubsection{Mersenne twister}\label{PRNGMT}
One of the most ``fashionable'' modern PRNGs is the Twisted GFSR
generator ({\it TGFSR}) or Mersenne twister generator
\cite{Matsumoto:1992tw,Matsumoto:1998mt}. \PRNG{TGFSR} is the
modernization of the GFSR and its algorithm is based on the
following recurrence for $w$-bit vectors $X_n$
\begin{eqnarray}
X_{n+r}^{\PRNG{TGFSR}}=X_{n+s}\XOR \left(X_{n}^{\rm upper}\OR
X_{n+1}^{\rm lower}\right)A,
\end{eqnarray}
where superscript indices ``upper'' and ``lower'' denote the $w-u$
highest and $u$ lowest bits of a corresponding binary vector
$X_i$, respectively,
\begin{eqnarray}
 X_{i}^{upper}&=&X_{i}\AND (2^{w}-2^{u}),\\\nonumber
 X_{i}^{lower}&=&X_{i}\AND (2^{u}-1),
\end{eqnarray}
and matrix $A$ is a ``twisting'' binary $w\times w$ matrix, which
form is chosen by performance reason
\begin{eqnarray}
A=\left(
\begin{array}{ccccc}
0 & 1 & 0 &  & 0\\
0 & 0 & 1 &  & 0\\
 &  &  & \ddots & \\
0 & 0 & 0 &  & 1\\
a_{w-1} & a_{w-2} & a_{w-3} & \cdots & a_0
\end{array}
\right).
\end{eqnarray}
So, only one $w$-bit vector $a=(a_{w-1},a_{w-2},\cdots,a_0)$
defines the product $X_{i}A$,
\begin{eqnarray}
X_{i}A=\Biggl\{
\begin{array}{ll}
(X_{i}\gg 1) & {\rm if~}(X_{i}\AND 1)=0\\
(X_{i}\gg 1)\XOR a & {\rm if~}(X_{i}\AND 1)=1
\end{array}.
\end{eqnarray}
For improving the statistical properties of the sequence so-called
tempering procedure is applied for the output sequence $X_n$. This
procedure is defined with the
\begin{eqnarray}
 t=X_{n}\XOR (X_{n}\gg m),\\\nonumber
 t=t\XOR ((t\ll d)\AND b),\\\nonumber
 t=t\XOR ((t\ll e)\AND c),\\\nonumber
 Y_{n}^{\PRNG{TGFSR}}=t\XOR(t \gg l).
\end{eqnarray}
By appropriate choosing of $r$, $u$ parameters and binary vector
$a$, it might reach the maximal period of the \PRNG{TGFSR},
\begin{eqnarray}
P_{\PRNG{TGFSR}}\leq (2^{rw-u}-1).
\end{eqnarray}

The most famous implementation of the \PRNG{TGFSR} PRNG is
\PRNG{MT19937} \cite{Matsumoto:1998mt}. The \PRNG{TGFSR}
parameters of the \PRNG{MT19937} are the following:
\begin{eqnarray}
 w=32,~r=624,~s=397,~u=31,\\\nonumber
 a={\rm 9908B0DF}_{16}=10011001000010001011000011011111_{2}
\end{eqnarray}
and tempering parameters are
\begin{eqnarray}
 m=11,~l=18,\\\nonumber
 d=7,~b={\rm 9D2C5680}_{16},\\\nonumber
 e=15,~c={\rm EFC60000}_{16}.
\end{eqnarray}
The maximal period of the \PRNG{MT19937} is
\begin{eqnarray}
 P_{\PRNG{MT19937}}\leq
2^{624\times 32-31}-1=2^{19937}-1\approx 4.3\times 10^{6001}.
\end{eqnarray}

\subsubsection{RANMAR}\label{PRNGRANMAR}
\PRNG{RANMAR} \cite{James:1988vf,Marsaglia:1987rm} is a
combination of two generators, 24-bit lagged Fibonacci generator
$LFG(97,33,-)$ $Y_n$ with $m=2^{24}=16777216$ and simple
arithmetic sequence $C_n$ for the prime modulus $M = 2^{24} - 3 =
16777213$,
\begin{eqnarray}
 X_n^{\PRNG{RANMAR}}=\left(Y_n-C_n\right)\mod m.
\end{eqnarray}
Or equivalently the producing recurrence is,
\begin{eqnarray}
 X_n^{\PRNG{RANMAR}}=Y_n-C_n,\\\nonumber
 X_n=X_n+m~~ {\rm for}~X_n<0.
\end{eqnarray}
Here $Y_n$ and $C_n$ are
\begin{eqnarray}
 Y_n=\left(Y_{n-97}-Y_{n-33}\right)\mod m,\\\nonumber
 C_n=\left(C_{n-1}-D\right)\mod M,
\end{eqnarray}
where $D=7654321$.

The $LFG(97,33,-)$ period is $P_{\PRNG{LFG(97,33,-)}}\leq
(2^{97}-1)2^{24}/2\simeq 2^{120}$ (see eqn.(\ref{LFGPeriods})) and
the period of the arithmetic sequence $C_n$ is
$P_{\PRNG{LCG(1,2^{24})}}\leq (2^{24}-1)$ (see
eqn.(\ref{LCGPeriod})). Therefore, the total period of the
\PRNG{RANMAR} is
\begin{eqnarray}
 P_{\PRNG{RANMAR}}\lesssim 2^{144}\simeq 2.23\times 10^{43}.
\end{eqnarray}

\subsubsection{Add-With-Carry and Subtract-With-Borrow generators}\label{PRNGAWC}
In 1991 Marsaglia and Zaman introduced a new class of PRNGs:
add-with-carry ({\it AWC}) and subtract-with-borrow ({\it SWB})
\cite{Marsaglia:1991nc}, which are small modifications of the LFG
with respect to supplementing an extra carry or borrow bit. Due to
the branching it became the first class of nonlinear PRNGs
\cite{Gentle:2003}. The AWC generator is described by the sequence
\begin{eqnarray}
X_n^{\PRNG{AWC}}=\left(X_{n-r}+X_{n-s}+c_{n-1}\right)\mod
m,\\\nonumber
 c_n=\Biggl\{
 \begin{array}{l}
 1~~ {\rm if}~ \left(X_{n-r}+X_{n-s}+c_{n-1}\right)\geq m\\
 0~~ {\rm if}~ \left(X_{n-r}+X_{n-s}+c_{n-1}\right)< m\\
 \end{array}.
\end{eqnarray}
In addition to $r$ seed values $(X_1,\ldots, X_r)$ generator must
be initialized with the carry bit $c_r$. The maximal period of the
AWC generator is
\begin{eqnarray}
P_{\PRNG{AWC}}\leq m^r+m^s-2.
\end{eqnarray}

In the SWB case the subsequent values of the sequence are obtained
by
\begin{eqnarray}\label{SWB}
X_n^{\PRNG{SWB}}=\left(X_{n-r}-X_{n-s}-c_{n-1}\right)\mod
m,\\\nonumber
 c_n=\Biggl\{
 \begin{array}{l}
 1~~ {\rm if}~ \left(X_{n-r}-X_{n-s}-c_{n-1}\right)< 0\\
 0~~ {\rm if}~ \left(X_{n-r}-X_{n-s}-c_{n-1}\right)\geq 0\\
 \end{array}.
\end{eqnarray}
L'Ecuyer noted \cite{Lecuyer:1994ur} that SWB has a second
variant, in which indices $r$ and $s$ in (\ref{SWB}) are swapped.
The maximal period of the SWB generator is
\begin{eqnarray}\label{SWBPeriod}
P_{\PRNG{SWB}}\leq m^r-m^s-2.
\end{eqnarray}

The well-known example of the SWB is \PRNG{RCARRY}
\cite{James:1988vf} which underlies the \PRNG{RANLUX} PRNG. The
RCARRY parameters in (\ref{SWB}) are $m=2^{24}$, $r=24$ and
$s=10$,
\begin{eqnarray}\label{RCARRY}
X_n^{\PRNG{RCARRY}}=\left(X_{n-24}-X_{n-10}-c_{n-1}\right)\mod
2^{24},\\\nonumber
 c_n=\Biggl\{
 \begin{array}{l}
 1~~ {\rm if}~ \left(X_{n-24}-X_{n-10}-c_{n-1}\right)< 0\\
 0~~ {\rm if}~ \left(X_{n-24}-X_{n-10}-c_{n-1}\right)\geq 0\\
 \end{array}.
\end{eqnarray}
The \PRNG{RCARRY} period is about \cite{Marsaglia:1991nc}
\begin{eqnarray}
P_{\PRNG{RCARRY}}\leq
\left((2^{24})^{24}-(2^{24})^{10}-2\right)/48\simeq 1/3\times
2^{572}\simeq 5.15\times 10^{171}.
\end{eqnarray}
It is less than the maximal period of the SWB (\ref{SWBPeriod})
because modulus $m=2^{24}$ is not a prime number.

\subsubsection{RANLUX}\label{PRNGRANLUX}
Despite all advantages (extremely long period, portability and
good productivity) AWC and SWB generators suffer from some
statistical defects (a bad lattice structure), showed by L\"uscher
\cite{Luscher:1993dy}. To eliminate these lacks James
\cite{James:1993vv} implements the L\"uscher's
\cite{Luscher:1993dy} idea to modify the SWB generator
\PRNG{RCARRY}. In the new generator which was called \PRNG{RANLUX}
(for LUXury RANdom numbers \cite{James:1993vv}) after producing
$r=24$ pseudo-random values the $p-r$ following sequential values
are discarded. It might be used any values of $p$, but there are
five generally accepted levels of the luxury every of which has
its own value $p$ \cite{Luscher:1993dy,James:1993vv},
\begin{itemize}
    \item {\bf level 0} ($p=24$): complete equivalent to original
    \PRNG{RCARRY} generator, there are no discarding values
    \item {\bf level 1} ($p=48$): throws out 24 values after
    one generation cycle; considerable improvement in quality over
    \PRNG{RCARRY}, passes the gap test, but still fails spectral
    test
    \item {\bf level 2} ($p=97$): passes all known tests, but
    theoretically still defective
    \item {\bf level 3} ($p=223$): default level, any
    theoretically possible correlations have a very small chance
    of being observed
    \item {\bf level 4} ($p=389$): highest possible luxury, all 24
    bits of the mantissa are chaotic.
\end{itemize}
L\"uscher recommends \cite{Luscher:1993dy} to use a default value
$p=223$ and notes that employment of values $p>389$ is pointless.

\end{document}